\documentclass{aa}  

\usepackage{graphicx}
\usepackage{txfonts}
\usepackage{amsmath}
\usepackage{amssymb}
\usepackage{wasysym}
\usepackage{booktabs}
\usepackage{tabularx}
\usepackage[linkcolor=blue, citecolor=blue, colorlinks=true]{hyperref}
\begin{document}

  \title{Redistribution of CO at the location of the CO ice line in evolving gas and dust disks}
  \titlerunning{Redistribution of CO in evolving gas and dust disks}

  \author{Sebastian Markus Stammler \inst{1,2} \thanks{Member of the
International Max Planck Research School for Astronomy and Cosmic Physics at the
Heidelberg University. PUC-HD Graduate Student Exchange Fellow.}
            \and
          Tilman Birnstiel \inst{3,4}
            \and
          Olja Panić \inst{5,6} \thanks{Royal Society Dorothy Hodgkin Fellow.}
            \and
          Cornelis Petrus Dullemond \inst{1}
            \and
          Carsten Dominik \inst{7}
         }

  \institute{Heidelberg University, Center for Astronomy, Institute of
Theoretical Astrophysics, Albert-Ueberle-Straße 2, 69120 Heidelberg, Germany \\
             \email{stammler@uni-heidelberg.de}
                \and
             Instituto de Astrofísica, Facultad de Física, Pontificia Universidad Católica de Chile, Santiago, Chile
                \and
             Max Planck Institute for Astronomy, Königstuhl 17, 69117
Heidelberg, Germany
                \and
             Harvard-Smithsonian Center for Astrophysics, 60 Garden Street,
Cambridge, MA 02138, USA
                \and
             School of Physics and Astronomy, The University of Leeds, E. C. Stoner Building, Leeds LS2 9JT, UK
                \and
             Institute of Astronomy, University of Cambridge, Madingley Road, Cambridge, CB3 0HA, UK
                \and
             University of Amsterdam, Astronomical Institute Anton Pannekoek,
Postbus 94249, 1090 GE Amsterdam, The Netherlands
            }

\abstract
   {Ice lines are suggested to play a significant role in grain growth and
planetesimal formation in protoplanetary disks. Evaporation fronts directly
influence the gas and ice abundances of volatile species in
the disk and therefore the coagulation physics and efficiency and the chemical composition of the
resulting planetesimals.}
   {In this work, we investigate the influence of the existence of the
CO ice line on particle growth and on the distribution of CO in the disk.}
   {We include the possibility of tracking the CO content and/or other volatiles in particles and in
the gas in our existing dust coagulation and disk evolution model and present
a method for studying evaporation and condensation of CO using the Hertz-Knudsen
equation. Our model does not yet include fragmentation, which will be part of further investigations.}
   {We find no enhanced grain growth immediately outside the ice line where the
particle size is limited by radial drift. Instead, we
find a depletion of solid material inside the ice line, which is solely due to
evaporation of the CO. Such a depression inside the ice line may be observable and
may help to quantify the processes described in this work. Furthermore, we find
that the viscosity and diffusivity of the gas heavily influence the
re-distribution of vaporized CO at the ice line and can lead to an increase in
the CO abundance by up to a factor of a few in the region just inside the ice line. Depending on the strength of
the gaseous transport mechanisms, the position of the ice line in our model can change by up
to $\sim 10\,\mathrm{AU}$ and consequently, the temperature at that location can range from \mbox{21}
to \mbox{23 K}.}
   {}
             
  \date{}
   
  \keywords{Protoplanetary disks --
            Accretion, accretion disks --
            Diffusion --
            Methods: numerical
           }

  \maketitle
%

\section{Introduction}

Ice lines are locations in protoplanetary disks at which a transition between the
gaseous and the solid phase of an element or a molecular species occurs. Inwards
of the ice line, the volatile species is condensed in the form of ice. Outside
the ice line, the material exists in its solid form. Ice lines are of special
interest in planetary sciences since they are suggested to affect the formation
and composition of planetesimals \citep{2011ApJ...743L..16O}. \citet{2014ApJ...793....9A}
argue from the chemical composition of Uranus and Neptune that they may have
been formed at the carbon monoxide ice line.

Planetesimals are formed from dust and ice surrounding young stars in protoplanetary
disks. The dust particles collide and are held together by contact forces
forming ever larger bodies \citep[see, e.g.,][]{2008A&A...480..859B,
2010A&A...513A..79B}. However, experiments and simulations show that the
growth is limited by at least two processes. One is the inward drift of particles on
Keplerian orbits losing angular momentum to the pressure-supported gas orbiting
the star with sub-Keplerian velocities \citep{1977MNRAS.180...57W}. The drift
velocity is size-dependent. Therefore, particles affected by drift quickly fall
onto the central star. Another mechanism that prevents grain growth is
fragmentation. If the collision velocity is high enough, silicate particles tend
to bounce or fragment rather than sticking together \citep{2008ARA&A..46...21B}.
If particles grow to certain sizes (centimeter to decimeter), the streaming
instability can take over and form planetesimals by clumping and subsequent
gravitational collapse \citep[see, e.g.,][]{2007Natur.448.1022J, 2005ApJ...620..459Y,
2010ApJ...722.1437B, 2014A&A...572A..78D}.

How particles can grow to sizes for which the streaming
instability is efficient is unknown. Besides the fact that icy materials have higher
fragmentation velocities than pure silicate particles \citep{2009ApJ...702.1490W, 2015ApJ...798...34G},
and can therefore grow to larger sizes, several authors argue that ice lines
play an important role in the growth process
\citep[e.g.,][]{2008IAUS..249..293K, 2008A&A...480..859B}. Also, \citet{2013A&A...557L...4K}
proposed that fluffy ice aggregates may overcome the drift barrier.
\mbox{\citet{1988Icar...75..146S}} found, in their model, a significant solid material
enhancement at the ice line due to diffusive redistribution of vapor from the
inner disk and subsequent condensation on particles outside the ice line.
Closely related, \citet{2013A&A...552A.137R} proposed grain growth by
evaporation of inward-drifting particles and subsequent recondensation of
backwards-diffusing vapor. \citet{2004ApJ...614..490C} proposed a significant
vapor enhancement inside the ice line due to rapidly inward-drifting particles.

The goal of our work is to include the transport of volatile materials into the
grain growth and disk model of \citet{2010A&A...513A..79B} and to use this model
to investigate
the influence of possible particle density enhancements at ice lines on dust
coagulation and on the distribution of CO gas. We therefore developed a model to track the volatile content of
particles and the gas including the processes of coagulation, radial drift, and viscous accretion, as well as
CO evaporation and condensation at the ice line.

We focus, here, on CO ice near the CO ice line for several reasons. First,
we assume that the collisional physics of the particles is mostly determined by the
presence of water ice. If we further assume that the water ice content is always
well mixed within the particle, there will always be enough water ice at the particle's
surface such that the collisional physics does not change by crossing the CO ice line and evaporating
the CO \citep[but see also][]{2015arXiv151003556O}. \citet{2009ApJ...702.1490W} estimated the
fragmentation velocity of water ice particles to be $\sim 50\ \mathrm{m /
s}$. Such high collision velocities are not reached in our model. We can
therefore neglect fragmentation, since, even by total evaporation of the CO, there
is still enough water ice in the particles such that we can use the higher
fragmentation velocity of water ice. The growth of the particles is therefore solely
limited by the inward drift. But see also \citet{2015ApJ...798...34G}, who found
fragmentation velocities for water ice in the order of $10\ \mathrm{m / s}$. It might also be
the case that the water ice is not well mixed within the particle or that the particle is covered
by a layer of CO. In that case, the fragmentation velocity is that of CO ice, which might also
be lower than the fragmentation velocity of water ice. We aim to investigate the influence of
fragmentation in future works.

Another reason for looking at CO is that it has observable features \citep{2015ApJ...813..128Q}.
The CO ice line at temperatures of approximately 20 K is in the outer parts
of protoplanetary disks compared to other volatile species such as water ice. Furthermore,
the CO abundance is generally large enough to be well observed.

We investigate the radial distribution of dust particles and molecular species
in the gas and ice phases around the region of the ice line in a time dependent
manner by solving the Smoluchowski equation and the re-distribution of the CO
vapor originating from the evaporating particles.

In section \ref{sectionModel}, we describe our model of advection, diffusion,
grain growth, and evaporation/condensation. In section \ref{sectionResults}, we
show the influence of the ice line on the grain growth and the influence of the
advection and diffusion of the gas on the CO vapor distribution at the ice
line. In section \ref{sectionDiscussion}, we discuss our results and the
assumptions made in the model.
  
\section{Model}
\label{sectionModel}

We constructed a model for simulating the evolution of a viscously evolving
protoplanetary disk including dust coagulation, particle drift, condensation, and evaporation
of CO at the CO ice line. The model is one-dimensional and the densities
are vertically integrated, assuming that gas and dust are constantly in thermal equilibrium, throughout the model.
We model the surface densities of approximately 150 different dust
sizes $\Sigma_{\mathrm{dust},i}$ and of two gas species, $\Sigma_\mathrm{H_2}$
and $\Sigma_\mathrm{CO}$. Our model of CO transport is built upon the gas and dust model of
\citet{2010A&A...513A..79B}.

The gas disk is viscously evolving according the $\alpha$-disk model of
\citet{1973A&A....24..337S} while we study the diffusive behavior of CO by
choosing different values for the Schmidt number Sc, which is the ratio of
viscosity over diffusivity of the gas.

We model dust growth by solving the Smoluchowski equation. The dust species
themselves are subject to radial drift and gas drag. In contrast to previous
works, we use a two-dimensional scalar field for the particle distribution,
where one dimension is the silicate mass and the second dimension the CO ice content of the
particle. We therefore model the migration of CO both via drifting
particles and via diffusing gas and simulate the evaporation of CO at the CO
ice line by solving the Hertz-Knudsen equation.

\subsection{Evolution of the gas surface density}

We consider two molecular gas species in our model, $\mathrm{H}_2$ , which is the main gas
density constituent, and $\mathrm{CO}$, which is generally the second most abundant molecular
species in the interstellar medium (ISM). The surface density of the two species, $\Sigma_\mathrm{H_2}$ and $\Sigma_\mathrm{CO}$,
and therefore the total gas density, $\Sigma_\mathrm{gas}$, are directly related through
the continuity equation:
\begin{equation}
  \frac{\partial}{\partial t} \Sigma_\mathrm{gas} + \frac{1}{R}
\frac{\partial}{\partial R} \left( R \Sigma_\mathrm{gas} u_\mathrm{gas} \right)
= 0,
  \label{eqnGasCont}
\end{equation}
where $u_\mathrm{gas}$ is the radial gas velocity given by
\citet{1974MNRAS.168..603L};
\begin{equation}
  u_\mathrm{gas} = -\frac{3}{\Sigma_\mathrm{gas}\sqrt{R}}
\frac{\partial}{\partial R} \left( \Sigma_\mathrm{gas} \nu \sqrt{R}
\right),
  \label{eqnGasVel}
\end{equation}
where $\nu$ is the viscosity of the gas. \citet{1973A&A....24..337S}
parameterized the viscosity in their $\alpha$-disk model to account for an
unknown source of viscosity with
\begin{equation}
  \nu = \alpha c_\mathrm{s} H_\mathrm{P},
\end{equation}
where $c_\mathrm{s}$ is the sound speed,
$H_\mathrm{P}=c_\mathrm{s}/\Omega_\mathrm{K}$ the pressure scale
height of the disk, and $\Omega_\mathrm{K}$ the Keplerian frequency. The viscosity parameter $\alpha$ is
typically of the order of $10^{-2}$ to $10^{-4}$ \citep[e.g.,][]{2005ApJ...634.1353J}.
Equation (\ref{eqnGasCont}) has no source terms on the right hand side, because
we do not consider any infall of matter from the common envelope onto the disk.
Evaporation and condensation, which would also be a source or sink term
for $\Sigma_\mathrm{CO}$, are treated separately.

We assume that both gas species evolve separately without
interacting with one another. We then obtain a separate continuity equation for
each of the species $j$
\begin{equation}
  \frac{\partial}{\partial t} \Sigma_j = -\frac{3}{R}\frac{\partial}{\partial
R} \left[ \sqrt{R} \frac{\partial}{\partial R} \left( \Sigma_j \nu \sqrt{R} 
\right) \right].
  \label{eqnViscAccr}
\end{equation}
Following the calculations of \citet{2007A&A...471..833P}, this equation can be
further manipulated to
\begin{equation}
  \begin{split}
    \frac{\partial}{\partial t} \Sigma_j &= -\frac{3}{R}\frac{\partial}{\partial
R} \left[ \sqrt{R} \frac{\partial}{\partial R} \left(
\Sigma_\mathrm{gas} \frac{\Sigma_j}{\Sigma_\mathrm{gas}} \nu \sqrt{R} \right)
\right] \\
        &= -\frac{3}{R}\frac{\partial}{\partial R} \left[ \sqrt{R}
\frac{\Sigma_j}{\Sigma_\mathrm{gas}} \frac{\partial}{\partial R} \left( \Sigma_j
\nu \sqrt{R} \right) + R \Sigma_\mathrm{gas} \nu
\frac{\partial}{\partial R} \left( \frac{\Sigma_j}{\Sigma_\mathrm{gas}} \right)
\right],
  \end{split}
\end{equation}
where $\Sigma_\mathrm{gas}=\sum\limits_j \Sigma_j$ is the total gas surface density.
By substituting the gas velocity of equation (\ref{eqnGasVel}) in this equation,
we obtain
\begin{equation}
  \frac{\partial}{\partial t} \Sigma_j + \frac{1}{R} \frac{\partial}{\partial
R} \left( R \Sigma_j u_\mathrm{gas} \right) -\frac{3}{R}
\frac{\partial}{\partial R} \left[ R \Sigma_\mathrm{gas} \nu
\frac{\partial}{\partial R} \left( \frac{\Sigma_j}{\Sigma_\mathrm{gas}} \right)
\right] = 0.
  \label{eqnGasAdvStandard}
\end{equation}
This equation implies that each gas species is radially advected with the gas
velocity $u_\mathrm{gas}$ and has a diffusive term, which smears out concentrations.

By introducing the Schmidt number $\mathrm{Sc} = \nu/D,$ which is the ratio of
viscosity $\nu$ to diffusivity $D$ of a gas, one can disentangle advection and diffusion
\begin{equation}
  \frac{\partial}{\partial t} \Sigma_j + \frac{1}{R} \frac{\partial}{\partial
R} \left( R \Sigma_j u_\mathrm{gas} \right) -\frac{1}{R}
\frac{\partial}{\partial R} \left[ R \Sigma_\mathrm{gas} D
\frac{\partial}{\partial R} \left( \frac{\Sigma_j}{\Sigma_\mathrm{gas}} \right)
\right] = 0.
  \label{eqnGasAdv}
\end{equation}
A Schmidt number of $\mathrm{Sc} = 1/3$ reproduces equation
(\ref{eqnGasAdvStandard}). Physically, the Schmidt number is the ratio of momentum
transport to pure-mass transport. By changing Sc and keeping $\alpha$ constant,
we investigate the influence of the diffusivity $D$ on the CO distribution
in the disk.

\subsubsection{Vertical and radial structure of the gas disk}

We assume that the gas is always in thermal equilibrium and that the temperature
profile does not change over time. The vertical structure
of the gas density is then given by
\begin{equation}
  \rho_\mathrm{gas} \left( z \right) = \rho_\mathrm{gas} \left( z=0 \right) \ 
\exp \left[ - \frac{1}{2} \frac{z^2}{H_\mathrm{P}^2} \right],
  \label{eqnGasVertStruct}
\end{equation}
where $z$ is the height above the midplane. Since $\Sigma_\mathrm{gas}$ is the
$z$-integral of $\rho_\mathrm{gas}$ , it follows directly for the midplane
gas density that
\begin{equation}
  \rho_\mathrm{gas} \left( z=0 \right) =
\frac{\Sigma_\mathrm{gas}}{\sqrt{2\pi}\ H_\mathrm{P}}.
\end{equation}

Even though the model is one-dimensional, we still need the vertical structure
of the disk to calculate the densities at the midplane that are needed for the
coagulation as well as for the evaporation method.

In our fiducial model, we used the following
power law as initial surface density distribution:
\begin{equation}
  \Sigma_\mathrm{gas} \left( R \right) = 250 \frac{\mathrm{g}}{\mathrm{cm}^2}
\times \left( \frac{R}{1\ \mathrm{AU}} \right)^{-1}.
\end{equation}

\subsubsection{Temperature structure of the disk}

For the midplane temperature of the disk, we assume a simple power law;
\begin{equation}
  T \left( R \right) = 150\ \mathrm{K}\ \times\ \left( \frac{R}{1\
\mathrm{AU}} \right)^{-\frac{1}{2}}
,\end{equation}
that remains constant in time. The viscous heating rate, which is the energy that
is created by the viscous accreting of the gas, is proportional to the gas
density $\Sigma_\mathrm{gas}$ \citep{1994ApJ...421..640N} and therefore mainly
important in the inner part of the disk. Since we are interested in the region
of the CO ice line at 40 – 50 AU where the densities are low, this effect is
negligible.

\subsection{Radial evolution of the dust surface density}

The spatial evolution of the dust particles is strongly influenced by
interactions with the gas. Although it is possible, in real disks, that particles
become locally concentrated to gas-to-dust ratios of unity or less, especially at the disk's
midplane, and therefore
trigger streaming instabilities, we neglect this here, and assume that accumulations of
dust happen only at smaller radii \citep{2002ApJ...580..494Y} or on smaller
scales \citep{2005ApJ...620..459Y}.
In other words: the gas has an effect on the dust particles, but the dust particles
have no back-reaction on the gas in our model.

To describe the coupling of the particles to the gas, it is useful to
characterize the particles not by size but by their dimensionless
\emph{Stokes number} St. The Stokes number is defined as the ratio of the
stopping time $\tau_\mathrm{s}$ of the particles to the largest eddy's turn-over time
$\tau_\mathrm{ed  }$;
\begin{equation}
  \mathrm{St} = \frac{\tau_\mathrm{s}}{\tau_\mathrm{ed}}.
  \label{eqnStokesDef}
\end{equation}
The Stokes number can be used as a measure to describe the coupling between the
gas and the dust. Since this coupling depends not solely on the particle size
but, amongst other parameters, also on the gas density, it is convenient to use the Stokes
number to describe the equations of motion of the particles.

The stopping time $\tau_\mathrm{s}$ in equation (\ref{eqnStokesDef}) is the
ratio between a particle's momentum and the rate of its momentum change by the
drag force. It yields the time scale on which a particle adapts to the gas velocity.
\citet{1977MNRAS.180...57W} identified four different
regimes of the drag force. The important regime in our work is the \emph{Epstein
regime}, where the stopping time is given by
\begin{equation}
  \tau_\mathrm{s} = \frac{\rho_\mathrm{s}\ a}{\rho_\mathrm{gas}\ \bar{u}}
,\end{equation}
with the bulk density of the solids $\rho_\mathrm{s}$ and the particle size
$a$. $\bar{u} = c_\mathrm{s}\sqrt{8/\pi}$ is the mean thermal velocity of the
gas molecules. In protoplanetary disks, the \emph{Stokes I} drag regime is
important only in the inner part of the disk ($\lesssim 1$ AU). We
therefore only include the Epstein regime in this work.

As a first order approximation for the eddy turnover time, one can use
$\tau_\mathrm{ed} = 1/\Omega_\mathrm{K}$ \citep{2004ApJ...614..960S},
and obtain the Stokes number:
\begin{equation}
  \mathrm{St} = \frac{\rho_\mathrm{s}\ a}{\Sigma_\mathrm{gas}}\frac{\pi}{2}.
\end{equation}

The advection of the dust surface densities of the  different species can
now be described with a continuity equation as equation (\ref{eqnGasAdv})
\begin{equation}
  \frac{\partial}{\partial t} \Sigma_{\mathrm{dust},i} + \frac{1}{R}
\frac{\partial}{\partial R} \left( R \left[ \Sigma_{\mathrm{dust},i}  
u_{\mathrm{dust},i} - D_{\mathrm{dust},i} \Sigma_\mathrm{gas} \frac{\partial}{\partial R} 
\frac{\Sigma_{\mathrm{dust},i}}{\Sigma_\mathrm{gas}} \right] \right) = 0,
  \label{eqnDustCont}
\end{equation}
where $D_{\mathrm{dust},i}= \alpha c_\mathrm{s} H_\mathrm{P} / (1 + \mathrm{St}^2)$ is the dust diffusivity and 
$u_{\mathrm{dust},i}$ 
the radial velocity of dust species $i$ given by
\begin{equation}
  u_{\mathrm{dust},i} = \frac{u_\mathrm{gas}}{1+\mathrm{St}_i^2} -
\frac{2 u_\mathrm{P}}{\mathrm{St}_i + \mathrm{St}_i^{-1}}.
  \label{eqnDustVel}
\end{equation}
This velocity consists of two terms. The first term describes the dust
particles that are dragged along with the inwards accreting gas; it is most
effective for small particles $\left( \mathrm{St} \ll 1 \right)$ that are
well-coupled to the gas and insignificant for large particles $\left(
\mathrm{St} \gg 1 \right)$.
The second term describes radial drift due to a pressure gradient in the gas.
The gas feels an additional force due to its own pressure gradient. If the
pressure gradient is pointing inwards, then the gas is on a sub-Keplerian orbit.
The acceleration of the dust particles due to the pressure force, on the other hand,
is negligible because of their high bulk density compared to the gas. Therefore, the
particles try to orbit with a Keplerian velocity. Due to this velocity difference with the gas,
they feel a constant headwind, lose angular momentum, and drift in the direction
of the pressure gradient. The maximal drift velocity (for particles with
$\mathrm{St}=1$) is given by \citet{1977MNRAS.180...57W} as
\begin{equation}
  u_\mathrm{P} = - \frac{\frac{\partial}{\partial R} P_\mathrm{gas}}{2
\rho_\mathrm{gas} \Omega_\mathrm{K}}.
  \label{eqnDustVelP}
\end{equation}

\subsection{Vertical structure of the dust disk}

Since the dust particles
do not feel any pressure amongst themselves, we cannot give a simple
pressure-equilibrium expression for the particle scale heights as for the gas
scale height $H_\mathrm{P}$. The particle scale heights are given by the
equilibrium of vertical settling by gravity and turbulent stirring and therefore
depend on the particle's Stokes number. Therefore, every particle size has its
own scale height. We follow, here, \citet{2008A&A...480..859B} for the description
of the scale height of species $i$ as
\begin{equation}
  h_i = H_\mathrm{P} \cdot \min \left( 1, \sqrt{\frac{\alpha}{\min \left(
\mathrm{St}_i, \frac{1}{2} \right) \cdot \left( 1 + \mathrm{St}_i^2 \right) }}
\right),
  \label{eqnDustScaleHeight}
\end{equation}
which limits the dust scale height to be, at maximum, equal to the gas scale
height $H_\mathrm{P}$. From this equation, it can be seen that larger particles
$\left( \mathrm{St} \gg 1 \right)$ are settled to the midplane, while small
particles $\left( \mathrm{St} \ll 1 \right)$ follow the gas scale height. A
stronger turbulence (i.e., larger turbulent viscosity parameter $\alpha$), in
general, increases the scale height up to a maximum of $H_\mathrm{P}$.

We assume that the particles are distributed with Gaussian distributions just as
the gas in equation (\ref{eqnGasVertStruct}), but with the dust particle's
specific scale height $h_i$ instead, which allows us to analytically
integrate the Smoluchowski equation vertically.

\subsection{Dust coagulation}

We include dust growth in our model to study the CO redistribution in protoplanetary disks.
The dust carries the CO ice inwards through the CO ice line. 
As seen in equation (\ref{eqnDustVel}), the Stokes
number, and therefore the particle size, is important for the drift velocity of
the dust particles.

We therefore include hit-and-stick collisions of particles in our model. We
neglect fragmentation because the CO ice line is well outside of the water ice
line, meaning that, everywhere in our model, the particles have a significant
amount of water ice, which is also mixed within the particle in such a way that the
particle collision properties are always determined by the water ice. As explained above, we
therefore neglect fragmentation.

\subsubsection{Smoluchowski equation}

The growth of the dust particles with mass $m$ by hit-and-stick collisions
without fragmentation can be described by the Smoluchowski equation

\begin{equation}
  \begin{split}
    \frac{\partial}{\partial t} f \left( m \right) = & \frac{1}{2}
\int\limits_0^\infty \mathrm{d}m' \int\limits_0^\infty \mathrm{d}m'' f \left( m'
\right) f \left( m'' \right) K \left( m'; m'' \right) \times \\
            & \quad \times \delta \left( m' + m'' - m \right) \\
        & - f \left( m \right) \int\limits_0^\infty \mathrm{d}m' f \left( m'
\right) K \left( m; m' \right),
  \end{split}
  \label{eqnSmolu1D}
\end{equation}
where the collision kernel $K \left( m; m' \right) = \sigma_\mathrm{geo} \left(
m; m' \right) \Delta u \left( m; m' \right)$ is the product of the geometrical
cross section and the relative velocity of the colliding particles. Equation
(\ref{eqnSmolu1D}) is a partial differential equation for the time evolution of
a mass distribution of particles. The positive term on the right-hand side
counts collisions that lead to particles of mass $m$. The negative term removes
particles of mass $m$ that collided with other particles to create larger
particles.

To keep track of not only the mass of the particles, but also of
another property that represents the CO ice content of them, we add
another parameter $Q$ to our distribution. The following equations are general
for an arbitrary $Q$ that could represent anything from the volume of the
particle as in \citet{2009ApJ...707.1247O} to the electrical charge of the
particle. In the following sections
we identify $Q$ with the particle's ice fraction.
The two-dimensional Smoluchowski equation then reads
as follows
\begin{equation}
  \begin{split}
    \frac{\partial}{\partial t} f \left( m, Q \right) = & \frac{1}{2}
\int\limits_0^\infty \mathrm{d}m' \int\limits_0^\infty \mathrm{d}m''
\int\limits_0^\infty \mathrm{d}Q' \int\limits_0^\infty \mathrm{d}Q'' \times \\
            & \quad \times f \left( m', Q' \right) f \left( m'', Q'' \right)
\times \\
            & \quad \times K \left( m', Q'; m'', Q'' \right) \times \\
            & \quad \times \delta \left( m' + m'' - m \right) \\
            & \quad \times \delta \left( Q^\mathrm{new} \left( m', Q'; m'', Q''
\right) - Q \right) \\
        & - f \left( m, Q \right) \int\limits_0^\infty \mathrm{d}m'
\int\limits_0^\infty \mathrm{d}Q' f \left( m', Q' \right) \times \\
            & \quad \times K \left( m, Q; m' Q' \right),
  \end{split}
  \label{eqnSmolu2D}
\end{equation}
where $Q^\mathrm{new} \left( m', Q'; m'', Q'' \right)$ is the new $Q$-value of
the particle that was created in the collision.
In principle, this equation could be solved as for equation (\ref{eqnSmolu1D}),
but adding another parameter and therefore two additional integrals would be
computationally very expensive.

We therefore follow the moment method introduced by \citet{2009ApJ...707.1247O} to
re-write one two-dimensional Smoluchowski equation into two one-dimensional
equations. The idea is that instead of evolving the full $Q$- distribution,
we only evolve its first moment, its mean $\bar{Q}$ for a given mass $m$.
By introducing the distribution function of particles of mass $m$
\begin{equation}
  n \left( m \right) = \int\limits_0^\infty f \left( m, Q \right) \mathrm{d}Q
,\end{equation}
and the mean $Q$-value of particles with mass $m$
\begin{equation}
  \bar{Q} \left( m \right) = \frac{1}{n \left( m \right)} \int\limits_0^\infty Q
f \left( m, Q \right) \mathrm{d}Q
,\end{equation}
and by assuming that the particle distribution is very sharp in $Q$
\begin{equation}
  f \left( m, Q \right) = n \left(m \right) \delta \left( \bar{Q} \left( m
\right) - Q \right),
\end{equation}
where $\delta$ is the Dirac delta function,
equation (\ref{eqnSmolu2D}) can be rewritten and simplified. We end up with one
partial differential equation for $n \left( m \right)$
\begin{equation}
  \begin{split}
    \frac{\partial}{\partial t} n \left( m \right) = & \frac{1}{2}
\int\limits_0^\infty \mathrm{d}m' n \left( m' \right) n \left( m - m' \right)
\times \\
            & \quad \times K \left( m', \bar{Q} \left( m' \right); m - m',
\bar{Q} \left( m - m' \right) \right) \\
        & - n \left( m \right) \int\limits_0^\infty \mathrm{d}m' n \left( m'
\right) K \left( m, \bar{Q} \left( m \right); m', \bar{Q} \left( m' \right)
\right)
  \end{split}
  \label{eqnSmoluN}
,\end{equation}
and one equation for $ n\bar{Q} \left( m \right) \equiv n \left( m \right)
\bar{Q} \left( m \right)$
\begin{equation}
  \begin{split}
    \frac{\partial}{\partial t} \left( n\bar{Q} \left( m \right) \right) =
& \frac{1}{2} \int\limits_0^\infty \mathrm{d}m' n \left( m' \right) n \left( m -
m' \right) \times \\
            & \quad \times K \left( m', \bar{Q} \left( m' \right); m - m',
\bar{Q} \left( m - m' \right) \right) \times \\
            & \quad \times Q^\mathrm{new} \left( m', \bar{Q} \left( m' \right);
m - m', \bar{Q} \left( m - m' \right) \right) \\
        & - n\bar{Q} \left( m \right) \int\limits_0^\infty \mathrm{d}m' n \left(
m' \right) K \left( m, \bar{Q} \left( m \right); m', \bar{Q} \left( m' \right)
\right).
  \end{split}
  \label{eqnSmoluNq}
\end{equation}
For a detailed derivation of the equations, we refer to
\citet{2009ApJ...707.1247O}. Both one-dimensional equations can be
simultaneously solved, which is significantly more efficient than solving one
two-dimensional equation.
The coagulation equations (\ref{eqnSmoluN}) and (\ref{eqnSmoluNq}) and the dust
advection equation (\ref{eqnGasAdv}) are simultaneously solved with the method
described in the appendix of \citet{2010A&A...513A..79B}

\subsubsection{The $Q$-parameter}

We have now the freedom of choice on what exactly the physical meaning of our $Q$-parameter should be.
One way to include ices in our coagulation code is to take
$m=m^\mathrm{tot}$ as the total mass and
$Q=\frac{m^\mathrm{ice}}{m}$ as the ice fraction of a particle. However, this approach
has the disadvantage that as a result of evaporation and condensation, both
parameters $m$ \emph{and} $Q$ change.
In this case, evaporation/condensation would be advection of the function $Q(m,t)$ 
in the $m$-coordinate with an additional source term.

We therefore chose $m=m^\mathrm{sil}$ as the silicate mass and
$Q=\frac{m^\mathrm{ice}}{m^\mathrm{sil}+m^\mathrm{ice}}$ as the ice fraction of a
particle. Here, evaporation and condensation only change $Q$ and  not $m$.

By using this approach, the ice mass of a particle is $m^\mathrm{ice} =
\frac{Q}{1-Q}m^\mathrm{sil}$  , whereas the total mass is $m^\mathrm{tot}
= \frac{1}{1-Q}m^\mathrm{sil}$. Therefore, we cannot
model particles fully consisting of ice $\left( Q = 1 \right)$.
$Q^\mathrm{new}$ , which is the new $Q$-value of a particle resulting from a
collision followed by sticking between a particle with mass $m$ and another particle with mass $m'$
(and ice fractions $Q$ and $Q'$, respectively), is then determined by
\begin{equation}
  \begin{split}
    Q^\mathrm{new} \left( m, Q; m', Q' \right) & = \frac{m^\mathrm{ice} +
{m^\mathrm{ice}}'}{m^\mathrm{tot}+{m^\mathrm{tot}}'} \\
        & = \frac{\frac{Q}{1-Q}m + \frac{Q'}{1-Q'}m'}{\frac{1}{1-Q}m +
\frac{1}{1-Q'}m'} \\
        & = \frac{ \left( 1 - Q' \right) Qm + \left( 1 - Q \right) Q'm' }{
\left( 1 - Q' \right) m + \left( 1 - Q \right) m' }.
  \end{split}
  \label{eqnQnew}
\end{equation}

\subsubsection{Relative velocities}

The coagulation is driven by the relative velocities of the particles that
determine the collision rates and the possible collision outcome. We consider,
here, five different sources of
relative velocities: Brownian motion $\Delta u_\mathrm{BM}$, azimuthal particle
drift $\Delta u_\phi$, radial particle drift $\Delta u_\mathrm{R}$, vertical
settling $\Delta u_\mathrm{sett}$  , and turbulence $\Delta u_\mathrm{turb}$.

The total relative velocity is then given by the root mean square
of all sources of relative velocity
\begin{equation}
  \Delta u = \sqrt{ \Delta u_\mathrm{BM}^2 + \Delta u_\mathrm{R}^2 + \Delta
u_\phi^2 + \Delta u_\mathrm{sett}^2 + \Delta u_\mathrm{turb}^2 }.
\end{equation}
We assume that all particles of a certain size collide with this distinct mean
velocity. It has been shown that the use of a velocity distribution can
have a significant influence on dust growth by breaking through growth barriers
as the bouncing barrier or the fragmentation barrier
\citep{2012A&A...544L..16W, 2013ApJ...764..146G}. Since we are not dealing
with bouncing and/or fragmentation in this model, we do not consider any
velocity distribution.

For a detailed description of the various sources of relative velocity, we refer to \citet{2010A&A...513A..79B}.

\subsection{Evaporation and condensation}
\label{sectionEvapCond}

When particles drift inwards they move from colder regions in the disk to warmer
regions. At some point – at the ice line – the temperature is high enough for the
ice in the particles to be evaporated. The vapor that is created by evaporation
can then be diffused backwards through the ice line and can re-condense on the
colder particles there. We therefore consider both evaporation and
condensation along with diffusion and viscous evolution of the gas in our
model.

We assume that the particles always adapt instantaneously to the new
temperature. This might not be the case for very large particles with high
heat capacities. But as seen later in Section \ref{sectionResults}, we do not
have such large particles. The temperature change our particles experience is solely
caused by their drift from colder regions to warmer regions in the disk. With typical
drift speeds, the particles experience a change in temperature of less than \mbox{1\,K} in \mbox{1000\,yrs}.
The assumption of instantaneous adaption to the ambient temperature is therefore well justified.

Evaporation/condensation can be described by the \emph{Hertz-Knudsen equation}
which gives the number of CO molecules $N$ per unit surface area that leave the
particle,
\begin{equation}
  \begin{split}
    \frac{\mathrm{d}}{\mathrm{d}t} N = v_\mathrm{therm}
\left( \frac{P^\mathrm{eq}}{k_\mathrm{B}T} - n_\mathrm{vap} \right),
  \end{split}
  \label{eqnHertzKnudsen}
\end{equation}
where $n_\mathrm{vap}$ is the
number density of vapor molecules, $P^\mathrm{eq}$ the saturation pressure
of the volatile species and $v_\mathrm{therm} = \sqrt{8k_\mathrm{B}T/\left( \pi
m_\mathrm{CO} \right)}$ is the mean thermal velocity of gas molecules of mass
$m_\mathrm{CO}$. If the ambient pressure of the volatile species equals the
saturation pressure then there is no net evaporation.

$P^\mathrm{eq}$ has to be determined experimentally. We use the values
from \citet{1985A&A...144..147L} as
\begin{equation}
  P^\mathrm{eq}_\mathrm{CO} \left( T \right) = 1
\frac{\mathrm{dyn}}{\mathrm{cm}^2} \times \exp \left[ -\frac{1030\
\mathrm{K}}{T} + 27.37 \right].
\end{equation}

Evaporation/condensation is, in general, dependent on the surface curvature of
the particle, where convex surfaces have slightly higher evaporation rates than
concave surfaces \citep{2011ApJ...735..131S}.
We do not consider those surface curvature effects in our model. This means if
equation (\ref{eqnHertzKnudsen}) is negative, we have condensation on all
particles independent of their size. If equation (\ref{eqnHertzKnudsen}) is
positive, all particles evaporate as long as they still carry some ice.

The Hertz-Knudsen equation needs to be solved for every particle size at every
radial and vertical position of the disk separately and simultaneously. But
since we are not dealing with the vertical structure of the disk in our model,
we need to transfer this expression into an expression for surface densities.
If we look at the midplane of the disk, the Hertz-Knudsen equation for the midplane vapor
mass density reads as
\begin{equation}
  \frac{\mathrm{d}}{\mathrm{d}t} \rho_\mathrm{vap} = \sum_i \pi a_i^2\ n_\mathrm{dust, i}\ v_\mathrm{therm}\ \left( 
\frac{m_\mathrm{CO} P^\mathrm{eq}}{k_\mathrm{B}T} - \rho_\mathrm{vap} \right),
  \label{eqnHertzKnudsenRho}
\end{equation}
where $a_i$ is the particle radius and $n_\mathrm{dust, i}$ the midplane number density of dust particles species $i$. This 
system 
is in equilibrium if
\begin{equation}
  \rho_\mathrm{vap} = \frac{m_\mathrm{CO}\ P^\mathrm{eq}}{k_\mathrm{B}T} \equiv \rho_\mathrm{vap}^\mathrm{eq},
\end{equation}
where $\rho_\mathrm{vap}^\mathrm{eq}$ is defined as the equilibrium vapor mass density.
If we assume that the vapor is always in pressure equilibrium, we get for the equilibrium vapor surface density
\begin{equation}
  \Sigma_\mathrm{vap}^\mathrm{eq} = \sqrt{ 2\pi }\ H_\mathrm{P}\ \rho_\mathrm{vap}^\mathrm{eq} = \sqrt{ 2\pi }\ H_\mathrm{P}\ 
\frac{m_\mathrm{CO}\ P^\mathrm{eq}}{k_\mathrm{B}T}.
  \label{eqnSigmaCoEq}
\end{equation}
Assuming this, we can transform equation (\ref{eqnHertzKnudsenRho}) into an expression for the vapor
surface density
\begin{equation}
  \begin{split}
    \frac{\mathrm{d}}{\mathrm{d}t} \Sigma_\mathrm{vap} &= \sum_i \pi a_i^2\ n_\mathrm{dust, i}\ v_\mathrm{therm}\ \left( 
\Sigma_\mathrm{vap}^\mathrm{eq} - \Sigma_\mathrm{vap} \right) \\
      &= \sum\limits_i \left( E_i - C_i \Sigma_\mathrm{vap} \right) \\
      &= E - C \Sigma_\mathrm{vap} \equiv -S,
      \label{eqnSigmaVapPart}
  \end{split}
\end{equation}
with
\begin{align}
  C_i &= \pi a_i^2 v_\mathrm{therm} n_{\mathrm{dust},i} \\
  E_i &= C_i \Sigma^\mathrm{eq}_\mathrm{vap}.
\end{align}

Therefore, the change in the ice surface density of particle
species $i$ is given by
\begin{equation}
  \frac{\partial}{\partial t} \Sigma_{\mathrm{ice},i} = C_i
\Sigma_\mathrm{vap} - E_i,
  \label{eqnSigmaIcePart}
\end{equation}
which is the difference between a condensation term $C_i
\Sigma_\mathrm{vap}$ and an evaporation term $E_i$. 
The deeper meaning behind this formalism is that we treat the
evaporation/condensation rates as if the particles were in the midplane, while
we assume that gas and particles are always in thermal equilibrium and
distributed according to their pressure scale heights $H_\mathrm{P}$ and $h_i$.

For $C_i$ and $E_i$ , the following relation holds:
\begin{equation}
  \frac{E_i}{C_i} = \frac{E}{C}.
  \label{eqnEC}
\end{equation}

\subsubsection{Condensation mode}

If $S>0,$ then we are in the \emph{condensation regime} and condensation will
take place on all particles. In that case, we can directly integrate equations
(\ref{eqnSigmaIcePart}) and (\ref{eqnSigmaVapPart}) until the end of the time
step. If we assume that the particle radius $a_i$ is constant and does not
change over the time step, then both equations have analytic solutions. This
assumption is justified if we use a canonical abundance of CO of $10^{-4}$. By
using a dust to gas ratio of $10^{-2}$ , this leads to an initial mass ratio of
CO (gas and ice phase) to dust of $\sim 14 \  \%,$ and therefore a maximal change in radius
of $\sim 5 \  \%$.
The analytic solution of equation (\ref{eqnSigmaVapPart}) is then
\begin{equation}
  \Sigma_\mathrm{vap} \left( t \right) = \Sigma_\mathrm{vap} \left( t_0 \right)
e^{-C\left( t-t_0 \right)} + \frac{E}{C} \left( 1-e^{-C\left( t-t_0 \right)}
\right).
  \label{eqnSigmaVapAnal}
\end{equation}
Using equation (\ref{eqnSigmaVapAnal}) in equation (\ref{eqnSigmaIcePart})
leads to
\begin{equation}
  \begin{split}
    \frac{\partial}{\partial t} \Sigma_{\mathrm{ice},i} = &-E_i + C_i
\Sigma_\mathrm{vap} \\
        = &-E_i + C_i \left( \Sigma_\mathrm{vap} \left( t_0 \right)
e^{-C\left( t-t_0 \right)} + \frac{E}{C} \left( 1-e^{-C\left( t-t_0 \right)}
\right) \right) \\
        = & \left( C_i \Sigma_\mathrm{vap} \left( t_0 \right) - E_i \right)
e^{-C\left( t-t_0 \right)},
  \end{split}
\end{equation}
where in the last step, equation (\ref{eqnEC}) was used. This equation has the
analytical solution
\begin{equation}
  \Sigma_{\mathrm{ice},i} \left( t \right) = \Sigma_{\mathrm{ice},i} \left( t_0
\right) + \frac{C_i \Sigma_\mathrm{vap} \left( t_0 \right)-E_i}{C} \left(
1-e^{-C\left( t-t_0 \right)} \right).
  \label{eqnSigmaIceAnal}
\end{equation}
Equations (\ref{eqnSigmaVapAnal}) and (\ref{eqnSigmaIceAnal}) can be used to
calculate the surface densities at any time step directly.

\subsubsection{Evaporation mode}

If at the start of the time step $S < 0$, we are in an \emph{evaporation
regime}. Also in this regime, the time evolution is governed by equations
(\ref{eqnSigmaVapAnal}) and (\ref{eqnSigmaIceAnal}). However, we need to take
into account that bare grains will not contribute to the flux of molecules, and
that during the time step, grains may become bare. Therefore, we split the
integration into a number of sub-steps. During each sub-step, the same set
of grain sizes contributes to the evaporation flux, and we completely ignore
bare grains. For each step, we define restricted versions of $C$ and $E$. These
restricted versions only sum over non-bare grains;
\begin{equation}
  C_\mathrm{rest} = \sum\limits_{i,\ \mathrm{not\ bare}}  C_i \qquad
\mathrm{and} \qquad E_\mathrm{rest} = \sum\limits_{i,\ \mathrm{not\ bare}} E_i.
\end{equation}
We define bare grains as particles that have less than one mono-layer of ice
molecules on their surface.

With these definitions, we use equation (\ref{eqnSigmaIceAnal}) to check if any
ice component becomes depleted during the time step, that is, we check for the first
root in any of the equations;
\begin{equation}
  \Sigma_{\mathrm{ice},i} \left( t_0 \right) + \frac{C_i \Sigma_\mathrm{vap}
\left( t_0 \right)-E_i}{C} \left( 1-e^{-C\left( t-t_0 \right)} \right) = 0.
  \label{eqnSigmaIceRoot}
\end{equation}
If that first root is before the end of the time step, we evolve the system to
this time, remove the newly bare grain size from the participating set of grain
sizes, and continue in the same way. Eventually, either all grains will be bare,
or the first root in the set of equations (\ref{eqnSigmaIceRoot}) will be beyond
the end of the time step. Then we use equation (\ref{eqnSigmaIceAnal}) with $C$
replaced by $C_\mathrm{rest}$ to move the system to the end of the time step.

\subsubsection{Caveats}

We do not consider grain curvature effects on evaporation and condensation. This
means either all particles are in evaporation mode or all particles are in
condensation mode. In general, this effect is mostly effective for very small
particles ($<\,0.1\,\mu$m). As seen in section \ref{sectionResults}, we get rid of the small particles
very quickly due to coagulation and do not replenish them since we do not have
fragmentation.

The surface of a real disk is directly illuminated by the star and therefore
hotter than the midplane. This means that in addition to a radial ice
line in the midplane, disks also have a two-dimensional surface called atmospheric
ice line towards their surface, where the temperatures are high enough for ices to evaporate.
Since we are only looking at midplane quantities to decide between
evaporation or condensation, we neglect the atmospheric ice line. This leads to
an overestimation of the ice content of the particles.
But since most of the dust mass is basically below the vertical
ice line, especially the larger grains,  the effect is
only expected to be important near the radial ice line where
the vertical ice line reaches the
midplane. However, this is a radially narrow region, so it shouldn't affect the global
transport, but possibly modifies the shape of the snow line.

Our particles have a relatively simple structure in the model. They are treated as
particles with a core and a mantle of CO. If two particles with mantles collide
in our model, they form a new particle with a single core and a mantle instead of
a particle
with two cores connected at their contact point, or even more complex fragmented structures.
This may lead to an artificial growth of silicate particles. If we consider the case
where outside the ice line, many particles with CO mantles
collide; in our model they form a large particle with a large single core and a
large CO mantle. By crossing the ice line and evaporating the CO, we are then
left with this large core instead of many small cores.
\citet{2011MNRAS.418L...1A} showed, experimentally, that by evaporating fractal
ice particles, one would expect to be left with many small particles instead of
one large particle. In favor of simplicity, we neglect this effect.

Furthermore, the evaporation of CO ice can be decreased if some of the CO ice is
trapped inside a porous and spatially complex structure of the dust particle.
Or it can be increased if the dust particle is very fluffy because its surface area
is then much larger than in the case of a spherical particle of the same size.
These effects are also ignored in this work.
  
\section{Results}
\label{sectionResults}

\subsection{The fiducial model}

\begin{table}
  \caption{Fiducial model parameters.}
  \begin{tabularx}{\hsize}{Xl l l l}
    \toprule \toprule
    Parameter & Symbol & Value & Unit \\
    \midrule
    Viscosity parameter & $\alpha$ & $10^{-3}$ & --  \\
    Schmidt number & $\mathrm{Sc}$ & $1/3$ & -- \\
    Stellar mass & $M_{\mathrm{star}}$ & 1.0 & $M_{\astrosun}$ \\
    Gas-to-dust ratio & $f_{\mathrm{g2d}}$ & 100 & -- \\
    Gas-to-CO ratio & $f_{\mathrm{g2CO}}$ & 700 & -- \\
    Silicate bulk density & $\rho_\mathrm{sil}$ & 1.6 & $\mathrm{g} / \mathrm{cm}^3$ \\
    CO bulk density & $\rho_\mathrm{CO}$ & 1.6 & $\mathrm{g} / \mathrm{cm}^3$ \\
    \bottomrule
  \end{tabularx}
  \label{tabInputs}
\end{table}

The input quantities of our fiducial model are given in table \ref{tabInputs}.
The grain sizes are initially distributed with the MRN distribution
\citep{1977ApJ...217..425M} with an upper cutoff at $1\ \mathrm{\mu m}$. We
integrate the equations (\ref{eqnGasAdv}), (\ref{eqnDustCont}),
(\ref{eqnSmoluN}), and (\ref{eqnSmoluNq}) and use the method for
evaporation/condensation as described in section (\ref{sectionEvapCond}) to calculate
the time evolution of the gas surface densities $\Sigma_{\mathrm{H}_2}$and
$\Sigma_\mathrm{CO}$, and the dust surface densities of the different grain sizes
$\Sigma_{\mathrm{dust},j}$.

\begin{figure}
  \resizebox{\hsize}{!}{\includegraphics{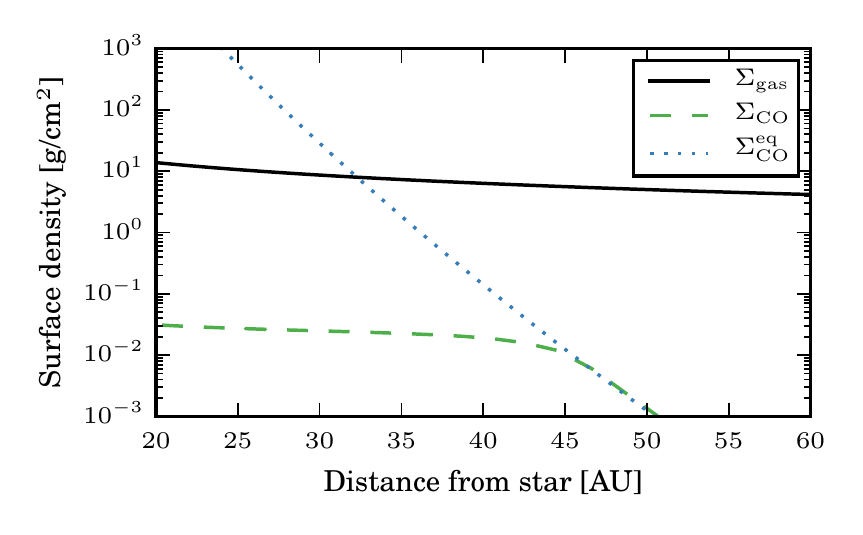}}
  \caption{Gas surface densities of the fiducial model after 1\,Myr. Shown are
the total gas surface density (\textit{solid black} line) and the surface density of
gaseous CO (\textit{dashed green} line). The \textit{dotted blue} line is the equilibrium
surface density of CO given by equation (\ref{eqnSigmaCoEq}). The location where
the CO surface density detaches from the equilibrium CO surface density is called
the \textit{ice line}.}
  \label{figFiducialGas}
\end{figure}

Figure \ref{figFiducialGas} shows the gas surface densities after 1\,Myr of
the simulation. We have chosen this value, because it is the typical age of
some of the best studied star-forming regions and thus also of the disks
therein. Also, it is the typical timescale for disk dispersion \citep[e.g.,][]{2001ApJ...553L.153H}.
Shown is the total gas surface density $\Sigma_\mathrm{gas} =
\Sigma_{\mathrm{H}_2} + \Sigma_\mathrm{CO}$ and the CO gas surface density
$\Sigma_\mathrm{CO}$. Overplotted is the equilibrium CO gas surface density
$\Sigma_\mathrm{CO}^\mathrm{eq}$ given by equation (\ref{eqnSigmaCoEq}), which the
system would have if there was no net evaporation/condensation and if there was
enough CO available. If there is not enough CO in the system, then the
CO in the particles evaporates until the grains are bare. The radial distance, where
$\Sigma_\mathrm{CO}$ detaches from $\Sigma_\mathrm{CO}^\mathrm{eq}$ , is then our
definition of the CO ice line because it is the point where there is just enough
CO in our model to provide the vapor pressure necessary to counterbalance the evaporation.
In our fiducial model, this happens at $\sim 47\ \mathrm{AU}$. Outside the ice
line, the CO is frozen out onto the grains until $\Sigma_\mathrm{CO} =
\Sigma_\mathrm{CO}^\mathrm{eq}$. Inside the ice line, all CO is in the gas phase.
Although it is not a power law, the equilibrium density at the position of the ice
line can be approximated by $\Sigma_\mathrm{CO}^\mathrm{eq} \propto
R^{-p}$ with $p \approx 20$. Since this is very steep, the region
around the ice line, where the CO is partially condensed, is very small and
therefore the ice line is well defined.

\begin{figure}
  \resizebox{\hsize}{!}{\includegraphics{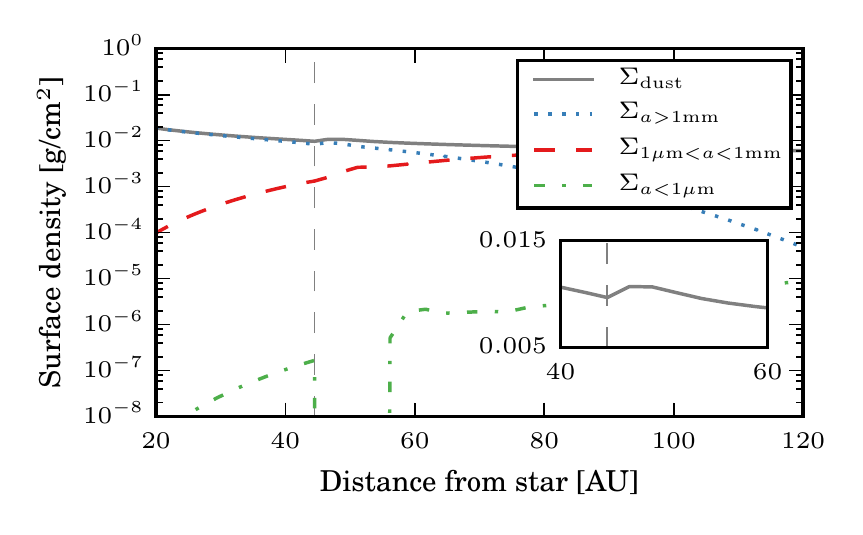}}
  \caption{Dust surface densities of different dust sizes at the CO
ice line after 1\,Myr. Shown are the total surface density ({solid gray} line),
millimeter-sized ({dotted blue} line), sub-millimeter-sized
(\textit{dashed red} line), and sub-micrometer-sized ({dash-dotted green} line)
particles. The vertical dashed line marks the location of the CO ice line.}
  \label{figFiducialDust}
\end{figure}

Figure \ref{figFiducialDust} shows the total dust surface density
$\Sigma_\mathrm{dust}$ and the surface
densities $\Sigma_{a>\mathrm{1mm}}$ of grains larger than one millimeter,
$\Sigma_{1 \mathrm{\mu m} < a < \mathrm{mm}}$ of grain sizes between 1 mm and 1 $\mu$m,
and $\Sigma_{a<\mathrm{1\mu m}}$ of sub-micrometer-sized grains after 1\,Myr.
The total dust surface density outside of 70 AU is
dominated by sub-millimeter-sized grains. The amount of millimeter-sized grains becomes
important inside of 70 AU. The reason for that
is that the maximum size of the particles is limited by radial drift. As
particles grow to larger sizes, their Stokes number increases. In the Epstein
regime of drag, which is relevant in this work, particles of the same size have a
larger Stokes number in regions of smaller gas density. This means that particles of
a given size drift faster in the outer region of the disk. When the particles
grow and their Stokes numbers approach unity, they are heavily affected by
radial drift and drift rapidly towards the central star.

We do not see any enhanced particle growth at the ice line. The reason for that
is the so-called drift barrier. If particles
drift through the ice line and evaporate their CO, this CO vapor can diffuse
backwards through the ice line and can re-condense on the particles there. These
particles then grow in size. Therefore, their Stokes number increases and with that their
drift velocity, making them drift even faster through the ice line.

In fact, the feature we see in $\Sigma_\mathrm{dust}$ in figure \ref{figFiducialDust}
at the ice line should rather be seen as a depletion inside the ice line instead of an
enhancement outside, because it is solely caused by evaporation of the CO ice on the inwards drifting
particles. This depletion approximately corresponds to the input ratios
of $f_\mathrm{g2d}$ and $f_\mathrm{g2CO}$.

Figure \ref{figFiducialDust} also shows a complete lack of sub-micrometer sized grains in a region
of approximately 10 AU outside the ice line. Those particles do not exist here, because the backwards
diffusing CO vapor preferentially re-condenses on the smallest particles since they contribute the most
to the total surface area. Therefore, these particles grow in size. The particles with the smallest
silicate cores have ice fractions of $Q>0.999$. Therefore, their mass is enhanced by a factor of
at least 1,000 and their radius by a factor of at least 10. Particles with a silicate core of
$a_\mathrm{sil}<1\,\mu m$ therefore have a resulting radius of $a_\mathrm{real}>1\,\mu m$.

\begin{figure}
  \resizebox{\hsize}{!}{\includegraphics{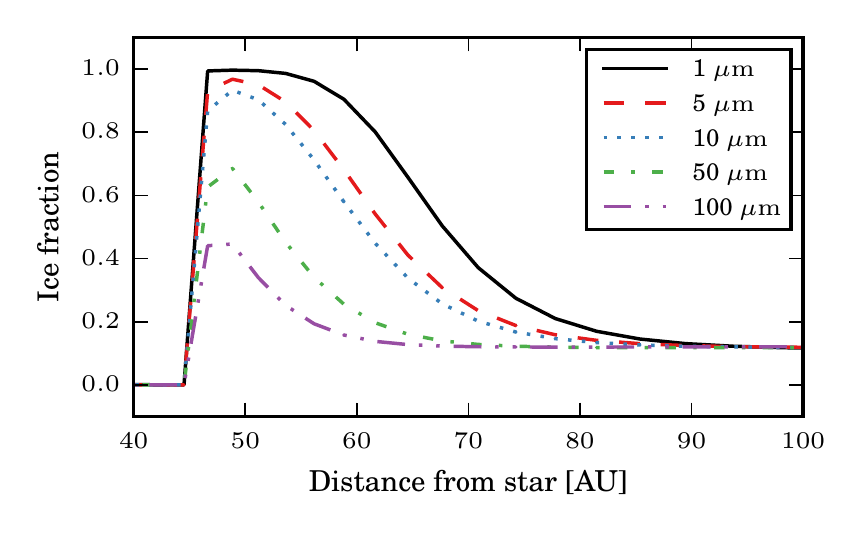}}
  \caption{The CO ice fraction $Q$ of particles with different silicate core sizes
depending on their radial position in the disk after 1\,Myr.}
  \label{figQofR}
\end{figure}

Figure \ref{figQofR} shows these ice fractions $Q\left(R\right)$ as a function of distance
from the star after 1\,Myr of the simulation for different silicate core sizes of the
particles. Since we do not take into account any curvature effects for our
evaporation/condensation algorithm, this means the evaporation/condensation
rate per unit surface area is the same for all particle sizes. In the case of
condensation, the absolute change in radius would be the same for all particles
independent of their size. But the relative gain in particle mass is therefore
larger for smaller particles and therefore their $Q$-value increases faster.
Figure \ref{figQofR} shows this effect. The smaller the particle is, the larger its CO ice fraction $Q$ at the ice line at 47\,AU.
Also, it can be seen that the CO vapor is redistributed to large distances
outwards of the ice line. Even at $R > 70\,\mathrm{AU,}$ there is a noticeable increase of $Q$
for the small particles.

\begin{figure*}
  \centering
  \includegraphics[width=17cm]{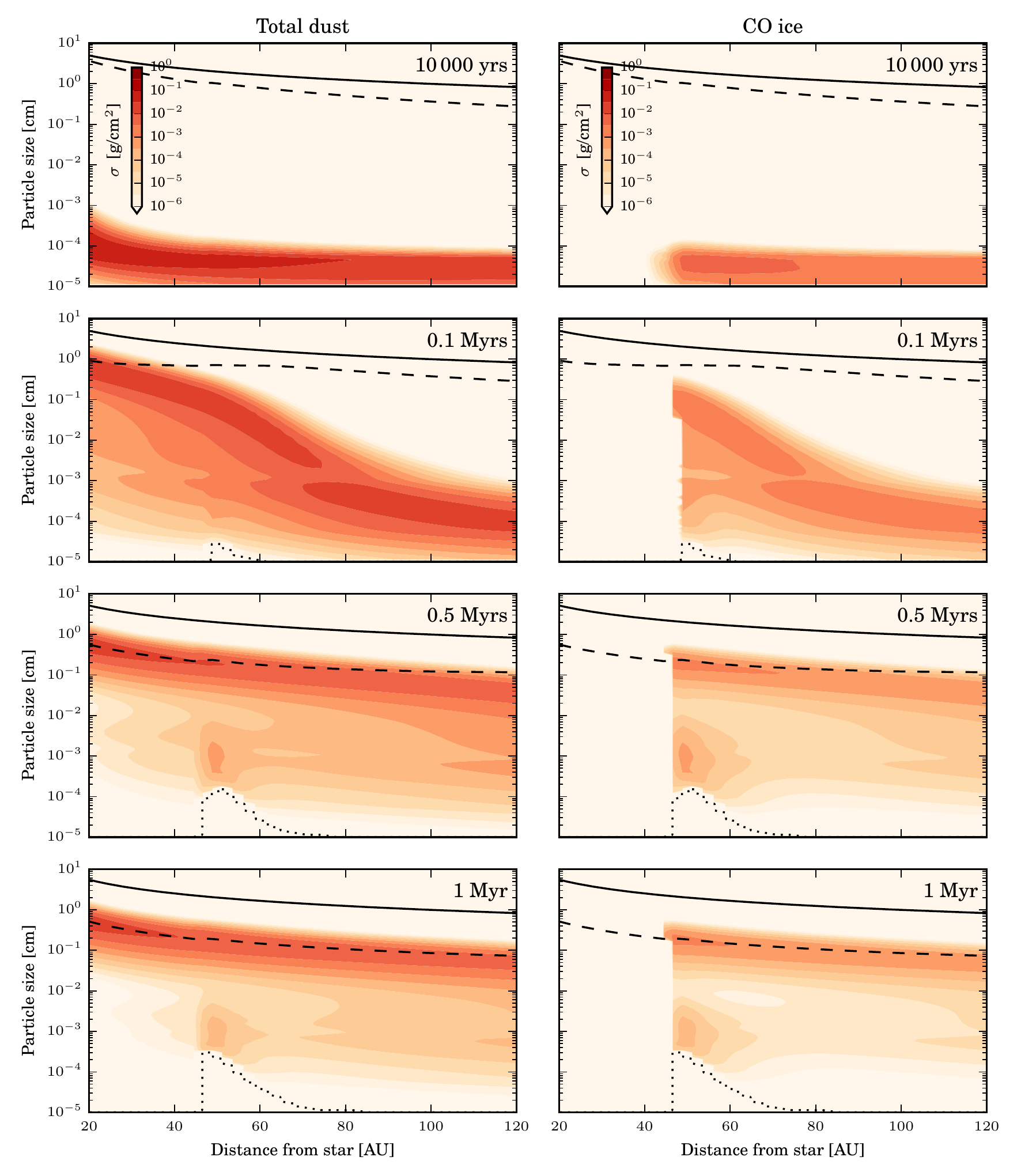}
  \caption{Different snapshots of our fiducial model. Shown are the different distribution functions as defined
  by equations (\ref{eqnSigmas}) at 10,000\,yrs, 100,000\,yrs, 500,000\,yrs, and at 1\,Myr. The solid line denotes
  the particle sizes with Stokes number unity, The dashed line is an analytical estimation of the drift barrier.
  The dotted line encompasses a region outside of the ice line that is free of small particles.}
  \label{figEvolution}
\end{figure*}

Figure \ref{figEvolution} shows different snapshots of the size evolution in our
fiducial model. Plotted are the following distribution functions:
\begin{equation}
  \begin{split}
    \sigma_\mathrm{dust,\,tot} \left( a; R \right) &= \int\limits_{-\infty}^{\infty} n \left( a; R, z \right) \cdot m \cdot 
a\;\mathrm{d}z \\
    \sigma_\mathrm{dust,\,CO} \left( a; R \right) &= \int\limits_{-\infty}^{\infty} nQ \left( a; R, z \right) \cdot m \cdot 
a\;\mathrm{d}z.
  \end{split}
  \label{eqnSigmas}
\end{equation}
The solid line denotes the particles with a Stokes number of unity. The dashed line
is an analytical estimate of the maximum drift-limited particle size as given by
\citet{2011A&A...525A..11B}.
At first, the particle growth is most effective in the inner regions
of the disk because of the larger collission rates of the particles. The particle
growth is later hindered by radial drift.
In the snapshots at 500,000\,yrs and 1\,Myr, a region from the ice line
at 47\,AU outwards can be seen where there are no small particles (dotted line). Because of the re-condensation
of CO, these particles grow in size. The distribution functions $\sigma$ are therefore compressed
towards larger sizes leading to the darker spot at intermediate particle sizes in the $\mu m$ range.
The total number density of particles has no discontinuity at the ice line.

\begin{figure}
  \resizebox{\hsize}{!}{\includegraphics{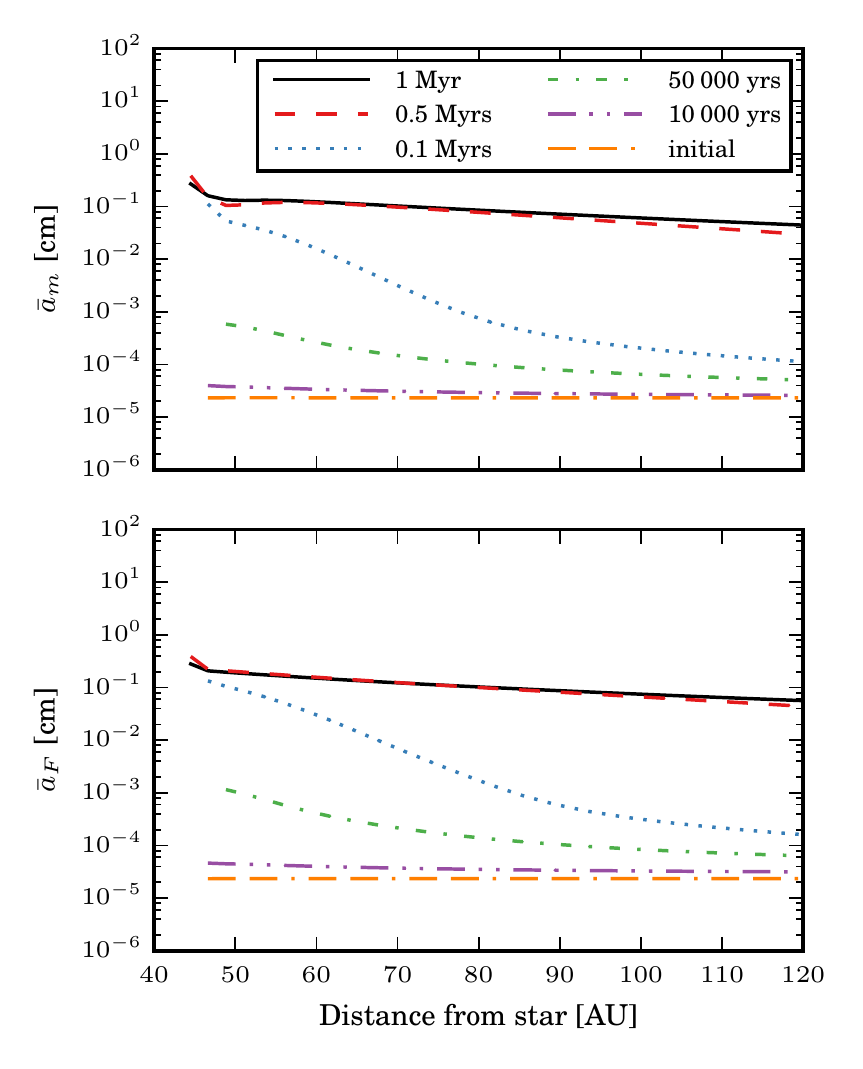}}
  \caption{The typical particle size $\bar{a}_m$ that carries most of the ice mass, and the typical
particle size $\bar{a}_F$ that transports most of the ice mass for different snapshots in our
fiducial model.}
  \label{figAverage}
\end{figure}

To investigate the typical particle size $\bar{a}_m$ that carries most of the CO ice mass
and the typical particle size $\bar{a}_F$ that transports most of the CO ice, we define the following quantities:
\begin{align}
  \bar{a}_m \left( R \right) &= \frac{ \int a \cdot \sigma_\mathrm{dust,\,CO} \, \mathrm{dln}a }{ \int \sigma_\mathrm{dust,\,CO} 
\, \mathrm{dln}a } \\
  \bar{a}_F \left( R \right) &= \frac{ \int a \cdot u_\mathrm{gas} \cdot \sigma_\mathrm{dust,\,CO} \, \mathrm{dln}a }{ \int 
u_\mathrm{gas} \cdot \sigma_\mathrm{dust,\,CO} \, \mathrm{dln}a }.
\end{align}
$\bar{a}_m$ is the average particle size weighted by the CO dust distribution and $\bar{a}_F$ is the average particle size
weighted by the CO ice flux. Both quantities are plotted in figure \ref{figAverage}. It can be seen by
comparing figure \ref{figAverage} with figure \ref{figEvolution} that $\bar{a}_m$ is always close to
the maximum particle size. Furthermore, $\bar{a}_m$ is always close to the particle size $\bar{a}_F$ that is responsible
for the ice flux on the particles in the disk.

\subsection{Transport of CO in the dust phase}

\begin{figure*}
  \centering
  \includegraphics[width=17cm]{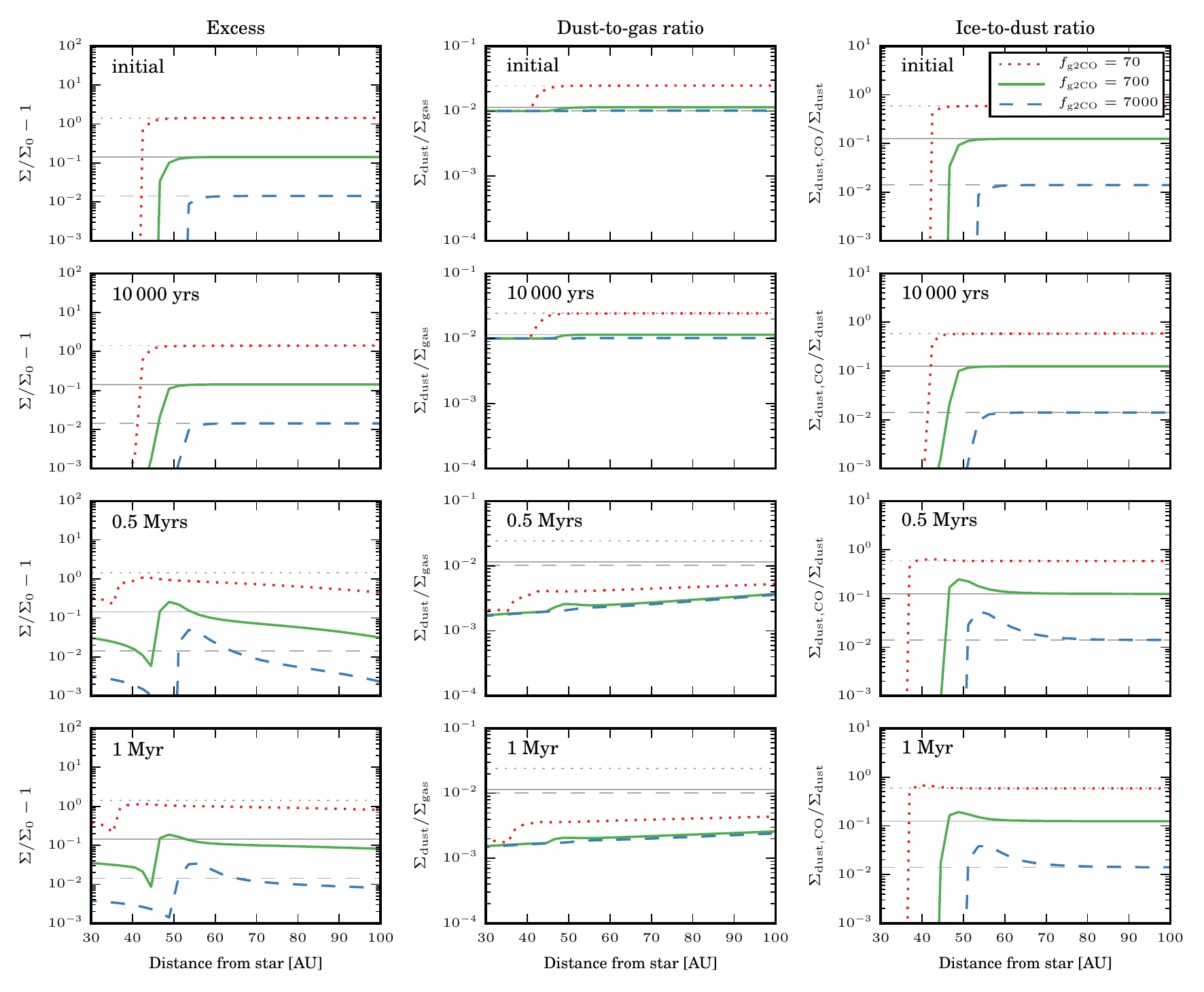}
  \caption{The excess of total dust surface density compared to a model without CO,
  the dust-to-gas ratio and the ice-to-dust ratio at different times for initial gas-to-CO
  ratios of 70 ({dotted red} line), 700 (fiducial model, {solid green} line)
  and 7000 ({dashed blue} line). The horizontal gray lines are the respective quantities
  one would expect from the initial conditions outside the ice line.}
  \label{figExcess}
\end{figure*}

Figure \ref{figExcess} shows the excess in total dust surface density compared to a model
without CO, the dust-to-gas ratio and the ice-to-dust ratio for different values of the
initial gas-to-CO ratio $f_\mathrm{g2CO}$ at different times in the simulation. We used values
for $f_\mathrm{g2CO}$ of 70, 700 (\textit{fiducial model}), and 7000, which correspond to
CO abundances of $10^{-3}$, $10^{-4}$, and $10^{-5}$. For guidance, the horizontal lines mark the values for
the quantities one would expect outside the ice line simply from the initial conditions.
The excess is defined as $\Sigma_i/\Sigma_0-1$, where $\Sigma_i$ is the total dust surface
density in the models of the different CO abundances and $\Sigma_0$ is the total dust surface
density in the model without CO. The initial value of the excess outside the ice line is
$f_\mathrm{g2d}/f_\mathrm{g2CO}$. The initial value of the gas-to-dust ratio is
$1/f_\mathrm{g2d}+1/f_\mathrm{g2CO}$. The ice-to-dust ratio compares the surface density of
CO ice to the total dust surface density. Its initial value is $f_\mathrm{g2d}/(f_\mathrm{g2d}+f_\mathrm{g2CO})$.

It can be seen that the closer the CO ice line is to the star, the higher the CO abundance is.
Further, the more CO there is in the system, the more saturated the evaporation/condensation is, causing the ice line to be
pushed inwards. This varies the position of the CO ice line in these models from approximately 40 AU to 50 AU, initially.

The left panels show the excess compared to a model without CO. Notably, the excess inside the ice line
is nonzero in all cases, meaning the total surface density is still
higher than in the comparison model without CO. Because of the larger CO-coated particles in the models with CO, and therefore the
more effective drift, the dust surface density is enhanced inside the ice line. This is especially visible in
the high abundance model.
In some models, the excess at the ice line can be a few times larger than one would expect from the initial
values. This is due to accumulation and re-distribution of CO at the ice line.

In general, the dust-to-gas ratio decreases with time. The dust surface density decreases while the
gas density remains approximately constant over time, as discussed later. The dust-to-gas ratio decreases from inside
out because the particle growth is most efficient in the inner parts of the disk, meaning the particles
start to drift earlier than outside. The dust-to-gas ratio is increased at the location of the ice
line. This is due to two effects: first, as the particles drift through the ice line, they shrink because of
evaporation, making drift less efficient and therefore leading to a ``traffic jam'', and second, the backwards-diffusing
and re-condensing CO vapor increases the ice surface density outside the ice line.

This effect can also be seen in the ice-to-dust ratio. At 1\,Myr, the ice-to-dust ratio is enhanced above
the initial value even at radii larger than 70\,AU depending on the model.

\subsection{Transport of CO in the gas phase}

\begin{figure}
  \resizebox{\hsize}{!}{\includegraphics{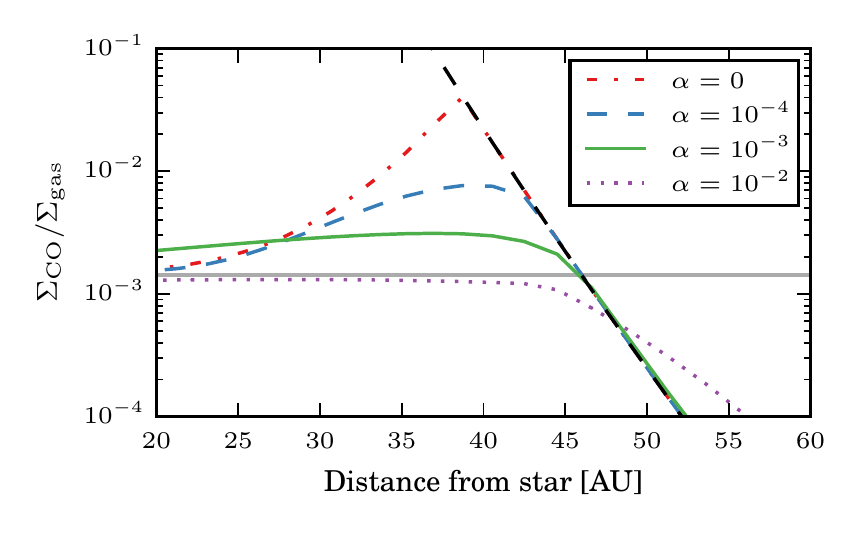}}
  \caption{The ratio of CO gas surface density to total gas surface density
  after 1\,Myr in the region of the ice line. Shown are the fiducial model
  with $\alpha=10^{-3}$ ({solid green line}),
a low-viscosity model with $\alpha=10^{-4}$ ({dashed blue}), and a high-viscosity model with $\alpha=10^{-2}$ ({dotted purple}). 
The ratio of
$\Sigma_\mathrm{CO}^\mathrm{eq}/\Sigma_\mathrm{gas}$ is
plotted with a dashed black line. Also plotted is the case
of a disk without viscous spreading ({red dash-dotted line}). The horizontal gray
line is our input value for the CO-to-gas mass ratio.}
  \label{figGasOverCOAlpha}
\end{figure}

CO is radially transported in the disk via three mechanisms: viscous accretion
and
diffusion in the gas phase and frozen-out as ice on drifting
particles. As, outside of the ice line, most of the CO is frozen-out on
particles, the dominant radial transport mechanism there is particle drift.
Inside the ice
line, on the other hand, all of the CO exists in the gas phase and is
radially transported by viscous accretion and diffusion.

This means that transport of CO through the ice line from the outside into the
inner part of the disk happens on drifting dust particles, while transport from
the
inside to the outside takes place in the gas phase. The efficiency of viscous
accretion is highly dependent on the viscosity parameter $\alpha$ while the
drift velocity of particles depends on their size and is only indirectly related
to $\alpha$ through the dust growth.

Particles are drifting through the ice line and deposit their CO as vapor there.
The strength of $\alpha$ determines how rapidly this CO vapor is diffused
from there. This is shown in figure \ref{figGasOverCOAlpha} where we compared
our fiducial model, with $\alpha = 10^{-3}$ , with a low-viscosity model of $\alpha
= 10^{-4}$ and a high-viscocity model of $\alpha = 10^{-2}$ at the time of 1\,Myr.
Plotted is the ratio of the CO gas surface density $\Sigma_\mathrm{CO}$ to the total gas surface density
$\Sigma_\mathrm{gas}$ at the region of the ice line. Also plotted is the ratio
of $\Sigma_\mathrm{CO}^\mathrm{eq}$ to $\Sigma_\mathrm{gas}$. The ice line is
therefore defined as the position where the CO gas
phase abundance approaches the level set by the equilibrium density (dashed
black line). This can happen anywhere from 39 to 49\,AU approximately, depending on the assumed
viscosity and consequently the CO gas abundance. The lower the viscosity, the more
efficient the accumulation of gas, pushing the ice line closer to the star.

Even though the temperature structure is the same for all three models, there
is still a difference in the location of the ice line.
The reason for this is that evaporation/condensation not only depends on
the temperature but also on the partial pressure of CO, as seen in equation
(\ref{eqnHertzKnudsen}). The partial pressure depends on the amount of
vaporized CO at a given location and on the temperature (cf. equation
(\ref{eqnSigmaCoEq})). This means that, in principle, any system can be in saturation (ie.,
equation (\ref{eqnHertzKnudsen}) equals zero) for any temperature if the density
is high enough. In our fiducial model, at 1\,Myr, (green solid line in figure
\ref{figGasOverCOAlpha}) this happens at \mbox{$\sim$ 47\,AU} where the
temperature is \mbox{$\sim$ 22 K}.

In the beginning of the simulations, the ice lines of the different models were
at the same position. As soon as the particles start to grow, they drift
inwards, cross the ice line, and deposit their CO there. Depending on the
efficiencies of accretion and mixing, the system can become saturated at the ice line if more CO
vapor gets deposited there by drifting particles than can be transported away
in the gas phase. If that is the case, then the ice line moves inwards because
the particles can only evaporate when they are in a region where the system
is not in saturation. Therefore, the ice line in the low-$\alpha$ models is closer
to the star than the ice line in the fiducial model. As mentioned above, this inward
motion of the ice line is entirely due to non-thermal effects, as we
consider a time invariant temperature structure.
Should the midplane temperature change, due to changes in disc flaring,  for example,
this would have a more noticeable effect on the ice line location as shown in \citet{2017MNRAS.tmp..115P}.

We also show, in Fig. \ref{figGasOverCOAlpha}, a model where we turned off the
viscous evolution of the gas disk by not solving equation (\ref{eqnGasAdv}). This is
equivalent to setting $\alpha=0$. In this case, all the CO
vapor remains where it is evaporated, and this moves the ice line inwards
to approximately 39\,AU compared to the fiducial case at 47\,AU.
In the region just inside the ice line, the ratio $\Sigma_\mathrm{CO}/\Sigma_\mathrm{gas}$
is largely increased compared to the initial ratio of $1/f_\mathrm{g2CO}$ depending
on the strength of viscosity. In the fiducial model, $\Sigma_\mathrm{CO}/\Sigma_\mathrm{gas}$
is, in general, slightly increased in the inner part of the disk, since the transport of evaporated CO
from the ice line to the inner disk is more effective here.

In the high viscosity case, the re-distribution of vapor is stronger than the recondensation onto
the particles. Meaning, vapor from inside the ice line becomes diffused outwards faster than it can recondense
onto the particles. This can be seen as the ratio of CO vapor to total gas is decoupled from
the equilibrium evaporation line. In this case, no increase of CO vapor in the inner disk can be seen.

\subsection{The influence of diffusion on the position of the ice line}

By rewriting the equation (\ref{eqnViscAccr}) for viscous accretion, one can see
that it consists of an advective and a diffusive term with a diffusivity $D =
3\nu,$ as can be seen in equation (\ref{eqnGasAdv}). By introducing the Schmidt
number $\mathrm{Sc}=\nu/D,$ we disentangled the advective and diffusive terms.
The Schmidt number compares the strength of angular momentum transport to pure
mass transport. To investigate the influence of the diffusivity on the transport
of CO, we performed simulations where we kept the viscosity, as in the fiducial
model, at $\alpha = 10^{-3}$ , and changed the diffusivity by choosing different
Schmidt numbers.

\begin{figure}
  \resizebox{\hsize}{!}{\includegraphics{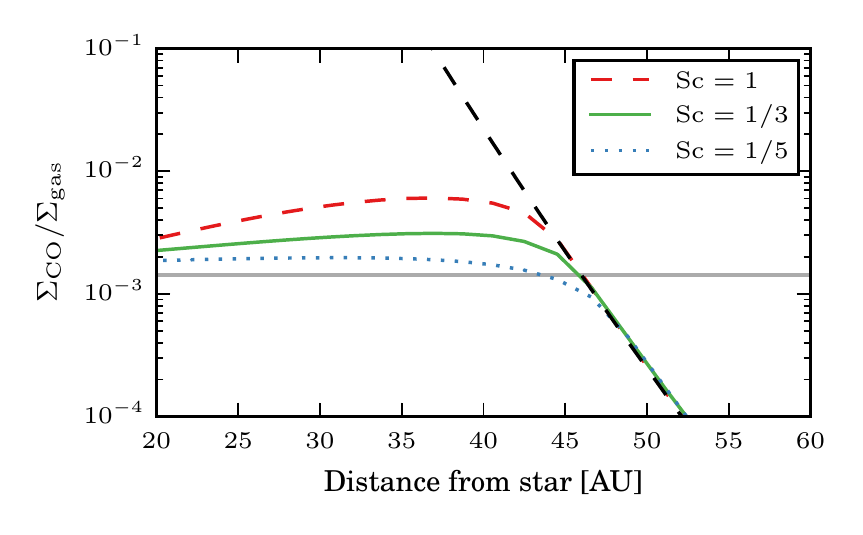}}
  \caption{Dependency of the ratio of CO vapor surface density to total gas
surface density on the Schmidt number at 1\,Myr. The fiducial model is
identical to the model with $\mathrm{Sc} = 1/3$ ({green solid line}). The dashed black line represents
the ratio of $\Sigma_\mathrm{CO}^\mathrm{eq}/\Sigma_\mathrm{gas}$. The
horizontal gray line corresponds to the initial CO-to-gas ratio.}
  \label{figSchmidtNumber}
\end{figure}

The result is shown in figure \ref{figSchmidtNumber}. The green models in
figures \ref{figGasOverCOAlpha} and \ref{figSchmidtNumber} are the fiducial
model and are therefore identical. It can be seen that the diffusivity $D$ is
relevant for distributing the CO in the gas disk; it smears out concentration
gradients. With lower diffusivities (higher Schmidt numbers), the pile-up of CO
gas just inside the ice line gets larger because it cannot be transported
rapidly enough as it gets deposited there. The ice line is therefore shifted
slightly to the inside of the disk for models with higher Schmidt numbers.

One possibility to trap particles is to assume a pressure bump. The dust
particles drift in the direction of the pressure gradient (cf. equation
(\ref{eqnDustVelP})). \citet{2007ApJ...664L..55K} proposed such a pressure bump
at ice lines because of a sharp transition of the turbulent viscosity parameter
$\alpha$ at the ice line. In our model, we do not change $\alpha$ at the ice line.
The pressure bumps we could create are therefore only due to evaporation of CO on
inward-drifting particles, but to create such a pressure bump only by depositing CO gas
is rather unrealistic since one would need to increase
the CO gas surface density by a factor larger than $f_\mathrm{g2CO}$.

\begin{figure}
  \resizebox{\hsize}{!}{\includegraphics{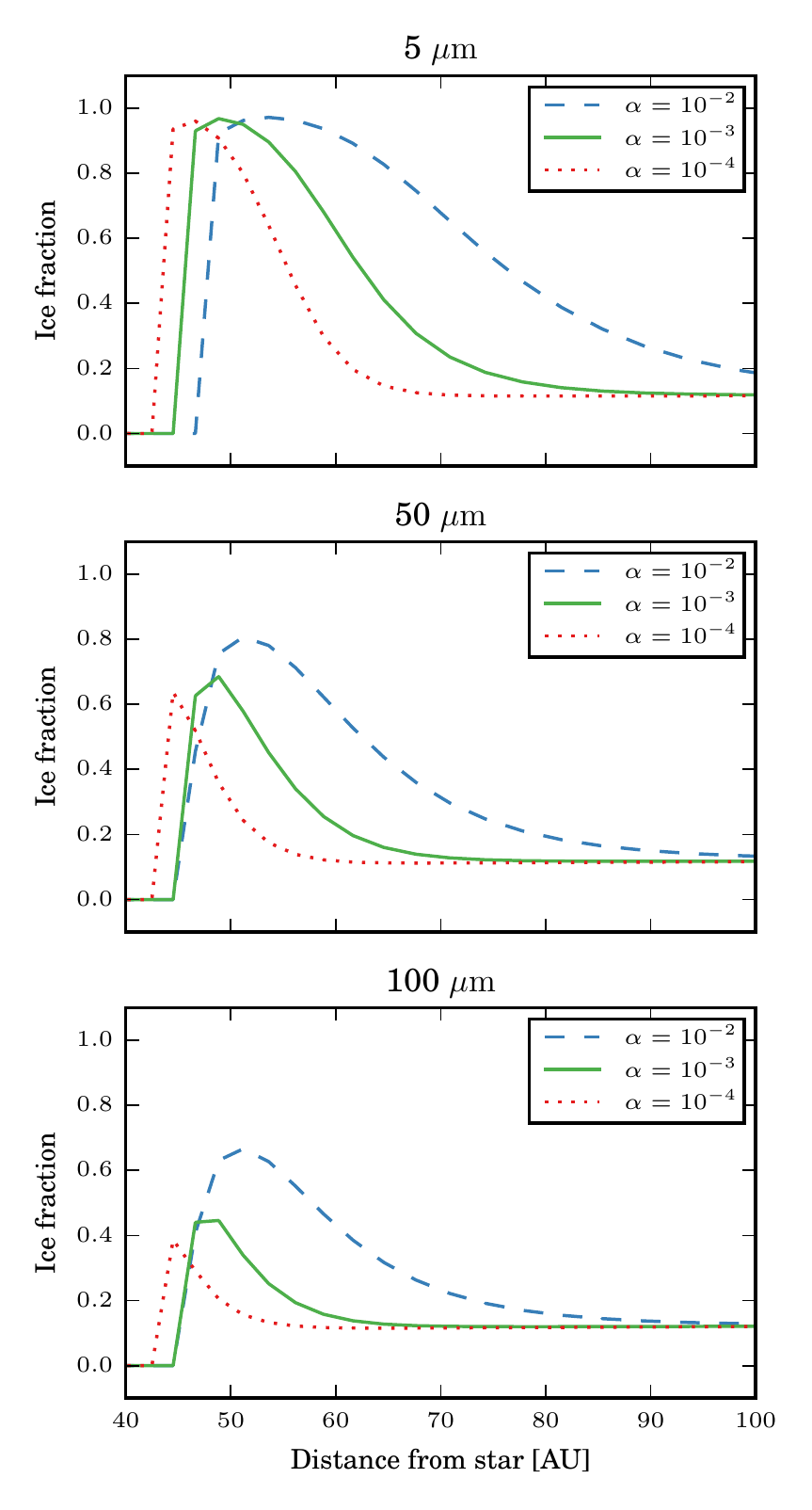}}
  \caption{The distribution of $Q$ according to  $R$ for different
  silicate core sizes $a_\mathrm{sil}$ and different values
  of $\alpha=10^{-4}$ ({dotted red} line), $\alpha=10^{-3}$ (fiducial, {solid green} line)
  and $\alpha=10^{-2}$ ({dashed blue} line) after 1\,Myr.}
  \label{figQofAlpha}
\end{figure}

Figure \ref{figQofAlpha} shows the distribution of ice along the different particle sizes
dependent on $\alpha$. For high-viscosity cases, the CO ice gets distributed as far out
as 90\,AU for small particles. In the low-viscosity case, the CO ice region is restricted up to 60\,AU. Also, the
higher the viscosity, the farther out the ice line. With high
$\alpha$ (and therefore, via the Schmidt number, also high $D$), CO vapor cannot be accumulated
at the ice line. This prevents any saturation effect.
  
\section{Discussion}
\label{sectionDiscussion}

\subsection{The CO surface density and ice line position}

In the previous section, we showed that $\Sigma_\mathrm{CO}$, the vapor density
of CO, approximately follows the equilibrium vapor density
$\Sigma_\mathrm{CO}^\mathrm{eq}$ , as long as there is enough CO in the disk.
If the evaporation/condensation timescales are shorter
than the dynamical time scales of the system, the CO  exactly follows the
equilibrium density; if the evaporation/condensation timescales are longer
than the dynamical timescales, it can decouple from the equilibrium vapor
density.
To first order, one can therefore approximate
the CO vapor surface density in the whole disk with the equation
\begin{equation}
  \Sigma_\mathrm{CO} = \min \left[ \frac{1}{f_\mathrm{g2CO}}
\Sigma_\mathrm{gas},\; \Sigma_\mathrm{CO}^\mathrm{eq} \right].
  \label{eqnSigmaCOApprox}
\end{equation}
Outside the ice line, the CO surface density is equal to
$\Sigma_\mathrm{gas}^\mathrm{CO}$; inside the ice line it is approximated
by using the total surface density and the initial gas-to-CO ratio. Although,
one should note that in figures \ref{figGasOverCOAlpha} and
\ref{figSchmidtNumber}, the ratio of $\Sigma_\mathrm{CO}/\Sigma_\mathrm{gas}$
can be significantly increased just inside the ice line, depending on the values
of $\alpha$ and Sc.

The position of the ice line $R_\mathrm{ice}$ is then defined as the point
where the CO surface density joins the equilibrium density and can be
approximated by equating:
\begin{equation}
  \frac{1}{f_\mathrm{g2CO}} \Sigma_\mathrm{gas} \left( R_\mathrm{ice} \right) =
\Sigma_\mathrm{CO}^\mathrm{eq} \left( R_\mathrm{ice} \right).
\end{equation}

\citet{2015ApJ...815..109P} showed, by comparing the desorption time $t_\mathrm{des}$
of evaporating particles with their respective drift timescale $t_\mathrm{drift}$, that
the location of the ice line does not necessarily coincide with the location where
the system gets out of saturation. When the particles drift faster than they evaporate
(i.e., $t_\mathrm{des} / t_\mathrm{drift} \gg 1$), the particles drift far past the
ice line before they evaporate their ice, leading to icy particles inside the ice line.

\begin{figure}
  \resizebox{\hsize}{!}{\includegraphics{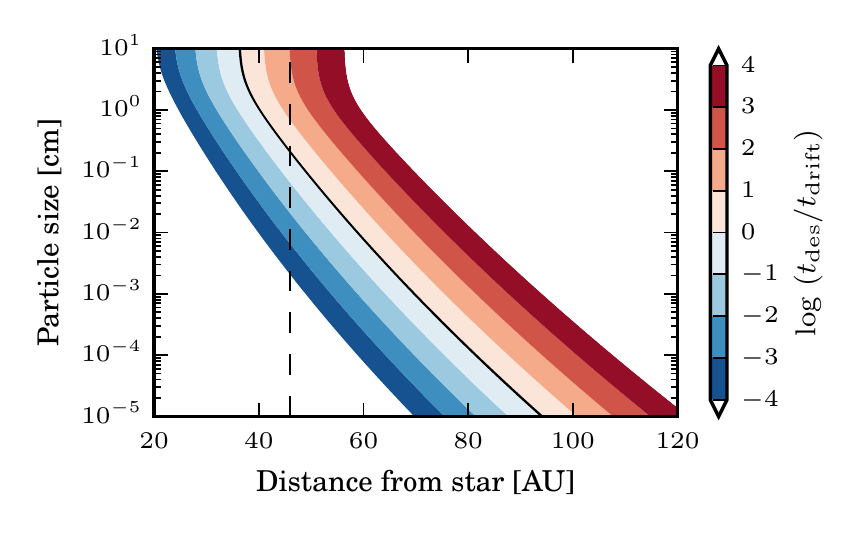}}
  \caption{Comparing the desorption time $t_\mathrm{des}$ with the drift timescale
  $t_\mathrm{drist}$. If the ratio $t_\mathrm{des}/t_\mathrm{drift}$ is larger than unity,
  the particles can drift past the ice line before they fully evaporate their ices. The vertical
  dashed line is the location of the ice line in our model.}
  \label{figTimescales}
\end{figure}

\begin{figure}
  \resizebox{\hsize}{!}{\includegraphics{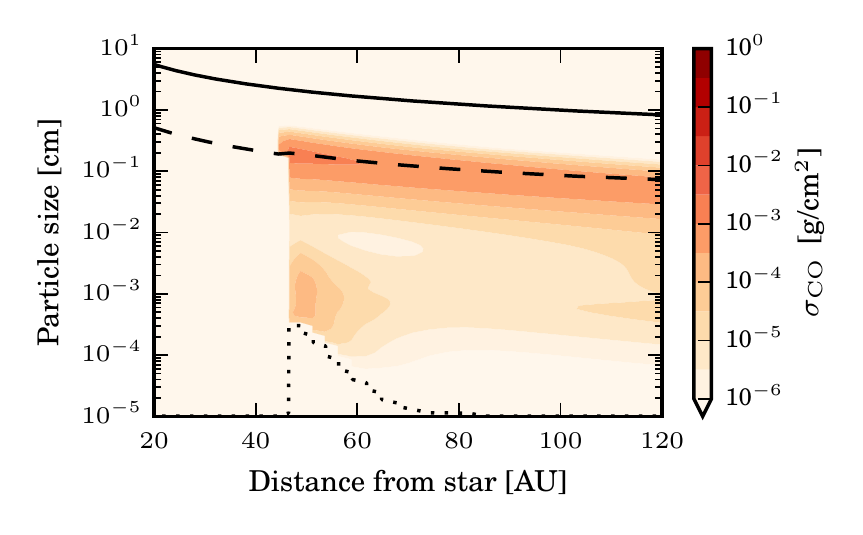}}
  \caption{CO ice surface density in our fiducial model after 1\,Myr.
  The largest particles are just large enough such that their drift timescale
  is smaller than their desorption timescale leading to a `smearing-out' of the ice line.
  The solid line denotes the particle sizes with Stokes number of unity. The dashed line is the drift barrier.
  The dotted line encompasses an empty region outside the ice line.}
  \label{figSmearing}
\end{figure}

Figure \ref{figTimescales} shows the ratio of the desorption time $t_\mathrm{des}$ to the
drift timescale $t_\mathrm{drift}$ for our fiducial model. The desorption time is given by
\begin{equation}
  t_\mathrm{des} = \frac{a}{\dot{a}} = \frac{a\,\rho_\mathrm{CO}}{m_\mathrm{CO}}\frac{k_\mathrm{B} T}{v_\mathrm{therm} 
P^\mathrm{eq}\left(T\right)},
\end{equation}
whereas the drift timescale is given by
\begin{equation}
  t_\mathrm{drift} = \frac{R}{u_\mathrm{dust}}.
\end{equation}
By comparing figure \ref{figTimescales} with figure \ref{figEvolution}, one can see that the
particles at the location of the ice line have a size where this `smearing-out' begins to
happen. Figure \ref{figSmearing} shows the CO ice surface density after 1\,Myr in our fiducial
model. The ice line of the smaller particles is at approximately 47\,AU, whereas the ice line for the
largest particles is at approximately 45\,AU.

By using equation (\ref{eqnSigmaCOApprox}), one has to keep in mind that the
values for $\Sigma_\mathrm{CO}$ around the ice line and the position of the ice
line itself highly depend on the transport properties of CO in the gas phase,
that is, the viscosity $\nu$ and therefore the $\alpha$ turbulence parameter, and on
the diffusivity $D$.
If the transport mechanism of CO in the gas phase is not strong enough, that is,
low $\alpha$ or low $D$, a pile-up of vaporized CO is created just inside
the ice line. This brings the system into saturation even closer to the star and
therefore shifts the ice line closer to the star. This can lead to differences
in the position of the ice line of up to 10\,AU and a change in temperature at the
ice line from 21 K to 23 K. In an evolving disk, where the
temperature profile is not fixed, the change in the position of the ice line can
be even stronger due to changes in the midplane temperature \citep[e.g.,][]{2017MNRAS.tmp..115P}.
Also, in the high viscosity case, redistribution of CO can be strong enough for the vapor
to be decoupled from the equilibrium vapor pressure.

By looking at figures \ref{figGasOverCOAlpha} and \ref{figSchmidtNumber}, one can
see that the CO vapor abundance in the inner part of the disk is significantly
increased over to the canonical ISM value. This was also predicted by
\citet[][Regimes 1 and 2 therein]{2004ApJ...614..490C}. However, we cannot reproduce
a depleted inner nebula as in \citet{1988Icar...75..146S} and \citet[][Regime 3]{2004ApJ...614..490C}
because we do not have any sink terms such as immobile planetesimals outside the ice
line that can remove the vapor from the disk. Our particles on which the CO vapor
re-condenses outside the ice line are still mobile and drift inwards through the ice line
replenishing the CO vapor there. Only in the high-viscosity case does the inner disk become slightly
depleted in CO vapor around the ice line due to the turbulent diffusion being effective enough
to push the gas farther out against recondensation.

\subsection{Particle growth at the ice line and detectability}

We do not see an enhanced particle growth just outside the ice line; instead we
see a depletion in surface density of dust inside the ice line. This depletion is entirely
due to loss of CO ice mantles, thereby making the dust particles smaller and less massive.
The particle surface density is dominated by
particles close to their local drift barrier. CO vapor, that previously diffused
backwards from inside the ice line and then recondensed on those particles
outside the ice line, only increase their size and Stokes number, making them
drift even more rapidly though the ice line. This drift barrier limits the
sizes of the particles in our models to less than one centimeter at the CO ice line.

The depression in total dust surface
density seen in figure \ref{figFiducialDust} is therefore related to
the amount of CO in the system. The radial jump in the total dust surface
density is approximately given by the ratio $\frac{f_\mathrm{g2d}}{f_\mathrm{g2CO}}$.
If we assume that the loss of the CO ice mantle at the CO ice line does not
cause any differences to the optical properties of the dust in the mm-wavelength regime, then the
difference in the intensity by thermal continuum emission of the dust $S_\nu$ is dominated by the
difference in surface density of the dust. Using the optically thin approximation
\begin{equation}
  \frac{S_{\nu,\mathrm{\ in}}}{S_{\nu,\mathrm{\ out}}} =
\frac{\Sigma_\mathrm{dust,\ in}}{\Sigma_\mathrm{dust,\ out}}
\frac{B_\nu(T_\mathrm{in})}{B_\nu(T_\mathrm{out})},
\end{equation}
where $B_\nu(T)$ is the Planck function and \textit{in}/\textit{out}
refers to locations shortly inside/outside the ice line. The dust densities
are given by
\begin{equation}
  \begin{aligned}
    \Sigma_\mathrm{dust,\ in} &=
\Sigma_\mathrm{dust}\left(R_\mathrm{in}\right)
\left(1-\frac{f_\mathrm{g2d}}{f_\mathrm{g2CO}}\right) \\
    \Sigma_\mathrm{dust,\ out} &=
\Sigma_\mathrm{dust}\left(R_\mathrm{out}\right) =
\Sigma_\mathrm{dust}\left(R_\mathrm{in}\right)
\left(\frac{R_\mathrm{out}}{R_\mathrm{in}}\right)^{p},
  \end{aligned}
\end{equation}
by assuming a simple power law for the dust surface density
$\Sigma_\mathrm{dust}\left(R\right)=\Sigma_0\left(\frac{R}{1\
\mathrm{AU}}\right)^{p}$.
This leads to
\begin{equation}
  \frac{S_{\nu,\mathrm{\ in}}}{S_{\nu,\mathrm{\ out}}} =
\left(1-\frac{f_\mathrm{g2d}}{f_\mathrm{g2CO}}\right)
\left(\frac{R_\mathrm{in}}{R_\mathrm{out}}\right)^{p}
\frac{B_\nu(T_\mathrm{in})}{B_\nu(T_\mathrm{out})} \simeq 0.86
\left(\frac{R_\mathrm{in}}{R_\mathrm{out}}\right)^{p},
\end{equation}
by assuming a canonical ISM value of $f_\mathrm{g2d}/f_\mathrm{g2CO}=1/7$ \citep[see e.g.,][]{1983A&A...122..171Y} and
$B_\nu(T_\mathrm{in})/B_\nu(T_\mathrm{out})\simeq1$. To detect the effect of
the CO evaporation and to quantify $f_\mathrm{g2d}/f_\mathrm{g2CO}$ (under the assumption
that no other process causes a bump at the ice line), it is clear
that one needs to constrain $p$ in the region beyond the ice line and measure
$S_\nu$ at $R_\mathrm{ice}$ to S/N much greater than 7.

\subsection{Influence of the particle size on the fragmentation velocity}

\citet{2015arXiv151003556O} showed that a change in the fragmentation velocity due to sintered
aggregates in the proximity of ice lines can produce rings in disks, as observed for example in
the disk around HL Tau. Evaporation preferentially happens on convex surfaces, while condensation
preferentially happens on concave surfaces. Close to the ice line, evaporation starts to happen
on the convex surface of a monomer, while simultaneously, condensation can happen on the concave contact
point (``neck'') of two connected monomers \citep{2011ApJ...735..131S}. This effect stiffens the neck
making it harder for the monomers to roll around this contact point. Aggregates can absorb the impact
energy of collisions by restructuring via rolling \citep{1997ApJ...480..647D}. Sintering reduces the
possibility of rolling leading to a breakup of the monomer connections and therefore a lower
fragmentation velocity.

In contrast to that, the critical breaking velocity necessary to break up the connection of two
monomers is inversely proportional to the monomer size $v_\mathrm{crit} \propto a_0^{-5/6}$
\citep[cf.][]{1997ApJ...480..647D} leading to lower fragmentation velocities for aggregates with
larger monomer sizes. We showed in figure \ref{figQofR} that our particles with a silicate core
of 0.1\,$\mu m$ (the monomers) have an ice fraction of $Q > 0.999$ in the region outside the ice line.
The monomer size in that region is therefore increased by a factor of at least 10 and the fragmentation
velocity decreased by a factor of approximately 7. This could lead to enhanced fragmentation just outside ice lines and
requires further investigation in upcoming works.

\subsection*{Possible improvements of the model}

\begin{enumerate}
  \item Our model is one-dimensional and we only use surface densities in our
calculations. This leads to errors in $\Sigma_\mathrm{CO}$ in our
treatment of the evaporation/condensation described in section
\ref{sectionEvapCond}. We underestimate $\Sigma_\mathrm{CO}$ since we are
neglecting the CO in the hot surface layers of the disk. However, since the disk's
atmosphere is small compared to its scale height, and its density is significantly smaller
than the midplane density, this should not be a large effect.
  \item We do not take into account the vertical movement of dust particles.
This movement could also influence the vertical distribution of CO
gas. One could imagine that the CO gas condenses on small dust particles in the
surface of the disk. These particles then sink more easily towards the midplane
and remove CO vapor from the disk's atmosphere. Or, vice versa, small CO-coated
dust particles are pushed through the atmospheric ice line due to turbulence,
depositing their CO as gas in the disk's surface. This is a topic of further
investigation.
  \item We also do not take into account the vertical diffusion of CO vapor and how
this affects the local CO vapor pressure in the midplane. As particles grow, they
settle to the midplane. When they drift inwards through the ice line, one would think
that most of the CO vapor is set free close to the midplane and is not instantaneously
in thermal equilibrium as we assume in our model.
  \item In our model we treat the particles as compact spheres, but
\citet{2013A&A...557L...4K} showed that, through fractal growth, the
particles can potentially overcome the drift barrier. The reason for this is
that fluffy aggregates have larger cross-sections than compact aggregates of the
same mass. This decreases the growth timescales of the particles, making the
growth of the particles more
efficient and potentially making it possible to overcome the drift barrier.
  \item Treating particles as compact spheres and not as fluffy aggregates
could lead to an artificial `non-fragmentation' by evaporation, since we combine the
silicate mass of the particle in one particle core after evaporation of the CO
instead of creating many small particles as \citet{2011MNRAS.418L...1A}
suggested.
  \item Including fragmentation and looking at the inner disk can potentially change some results. In general, the
particle size in the inner parts of the nebula is fragmentation limited instead of drift
limited, as in the outer disk. Therefore, the particle dynamics and distributions change, which
can have effects on the various volatile re-distribution mechanisms discussed in this paper.
\end{enumerate}

\section{Conclusions and outlook}
\label{sectionConclusions} 

We developed a model for dust growth with a volatile species including
particle drift and diffusion, evaporation and condensation of the volatiles, and
viscous evolution of the gas. We do not find any enhanced particle growth by re-condensation just outside the ice line,
since the particles there are already close to their drift barrier.
Due to particle drift through the ice line, evaporation, diffusion and re-condensation
on particles outside the ice line, we find a dust surface density enhancement at
the ice line that could be observable at the CO ice line.
The vapor that gets deposited at the ice line by inwards-drifting particles can
push the ice line closer to the star. If the efficiency of re-distribution of vapor in
the disk is low, then the vapor accumulates at the ice line, bringing the system into
saturation and therefore preventing evaporation.
The location of the ice line itself does depend on disk parameters such as temperature,
volatile abundance, viscosity, or diffusivity. Furthermore, different particle sizes
can have, in general, different locations at which they fully evaporate their volatile species.
The increase in monomer size just outside the ice line could lead to fragmentation,
exactly as proposed for sintering.

Our model is not restricted to CO. By using the analogous parameters for
water ice, we can transfer the simulation to the water ice line in the inner
disk. Here, one has to take into account that the collisional properties of
water-ice-coated particles and purely silicate particles differ tremendously
\citep[e.g.,][]{2008ARA&A..46...21B, 2010A&A...513A..56G}. For
ice-free particles, fragmentation plays an important role and has to be taken
into account for particles crossing the ice line. Furthermore, the water ice
line is at locations in the disk where viscous heating is important and
has to be taken into account. The question of the location and distribution of
$\mathrm{H}_2\mathrm{O}$ has an important role in astrophysics and planetary
sciences. Whether the Earth formed dry, that is, inside the water ice line, or in a region
outside the ice line remains to be determined. If the Earth formed dry, the origin of the Earth's water is unclear. If
it formed wet, the relatively small amount of water observed on Earth is curious. By
transforming our model to water ice and implementing fragmentation, we will try to
come closer to elucidating these subjects in future work.
 
\begin{acknowledgements}

S.M.S. has been supported by the Deutsche Forschungsgemeinschaft
Schwerpunktprogramm (DFG SPP 1385) ‘‘The first 10 Million Years of the Solar
System – a Planetary Materials Approach’’. 
S.M.S. gratefully acknowledges support through the PUC-HD Graduate Student Exchange
Fellowship, which is part of the academic exchange program between the Institute of Astrophysics
of the Pontificia Universidad Católica (IA-PUC) and the Center for Astrophysics at the University of Heidelberg (ZAH),
financed by the German Academic Exchange Service (DAAD).
The work of O.P. is supported by the European Union through the grant number 279973, and by the Royal Society Dorothy Hodgkin 
Fellowship, number 140243.

\end{acknowledgements}

  \bibliographystyle{elsarticle-harv}
  \bibliography{bibliography.bib}

\begin{thebibliography}{41}
\expandafter\ifx\csname natexlab\endcsname\relax\def\natexlab#1{#1}\fi
\expandafter\ifx\csname url\endcsname\relax
  \def\url#1{\texttt{#1}}\fi
\expandafter\ifx\csname urlprefix\endcsname\relax\def\urlprefix{URL }\fi

\bibitem[{{Ali-Dib} et~al.(2014){Ali-Dib}, {Mousis}, {Petit}, and
  {Lunine}}]{2014ApJ...793....9A}
{Ali-Dib}, M., {Mousis}, O., {Petit}, J.-M., {Lunine}, J.~I., Sep. 2014. {The
  Measured Compositions of Uranus and Neptune from their Formation on the CO
  Ice Line}. \apj 793, 9.

\bibitem[{{Aumatell} and {Wurm}(2011)}]{2011MNRAS.418L...1A}
{Aumatell}, G., {Wurm}, G., Nov. 2011. {Breaking the ice: planetesimal
  formation at the snowline}. \mnras 418, L1--L5.

\bibitem[{{Bai} and {Stone}(2010)}]{2010ApJ...722.1437B}
{Bai}, X.-N., {Stone}, J.~M., Oct. 2010. {Dynamics of Solids in the Midplane of
  Protoplanetary Disks: Implications for Planetesimal Formation}. \apj 722,
  1437--1459.

\bibitem[{{Birnstiel} et~al.(2010){Birnstiel}, {Dullemond}, and
  {Brauer}}]{2010A&A...513A..79B}
{Birnstiel}, T., {Dullemond}, C.~P., {Brauer}, F., Apr. 2010. {Gas- and dust
  evolution in protoplanetary disks}. \aap 513, A79.

\bibitem[{{Birnstiel} et~al.(2011){Birnstiel}, {Ormel}, and
  {Dullemond}}]{2011A&A...525A..11B}
{Birnstiel}, T., {Ormel}, C.~W., {Dullemond}, C.~P., Jan. 2011. {Dust size
  distributions in coagulation/fragmentation equilibrium: numerical solutions
  and analytical fits}. \aap 525, A11.

\bibitem[{{Blum} and {Wurm}(2008)}]{2008ARA&A..46...21B}
{Blum}, J., {Wurm}, G., Sep. 2008. {The Growth Mechanisms of Macroscopic Bodies
  in Protoplanetary Disks}. \araa 46, 21--56.

\bibitem[{{Brauer} et~al.(2008){Brauer}, {Dullemond}, and
  {Henning}}]{2008A&A...480..859B}
{Brauer}, F., {Dullemond}, C.~P., {Henning}, T., Mar. 2008. {Coagulation,
  fragmentation and radial motion of solid particles in protoplanetary disks}.
  \aap 480, 859--877.

\bibitem[{{Cuzzi} and {Zahnle}(2004)}]{2004ApJ...614..490C}
{Cuzzi}, J.~N., {Zahnle}, K.~J., Oct. 2004. {Material Enhancement in
  Protoplanetary Nebulae by Particle Drift through Evaporation Fronts}. \apj
  614, 490--496.

\bibitem[{{Dominik} and {Tielens}(1997)}]{1997ApJ...480..647D}
{Dominik}, C., {Tielens}, A.~G.~G.~M., May 1997. {The Physics of Dust
  Coagulation and the Structure of Dust Aggregates in Space}. \apj 480,
  647--673.

\bibitem[{{Dr{\c a}{\.z}kowska} and {Dullemond}(2014)}]{2014A&A...572A..78D}
{Dr{\c a}{\.z}kowska}, J., {Dullemond}, C.~P., Dec. 2014. {Can dust coagulation
  trigger streaming instability?} \aap 572, A78.

\bibitem[{{Garaud} et~al.(2013){Garaud}, {Meru}, {Galvagni}, and
  {Olczak}}]{2013ApJ...764..146G}
{Garaud}, P., {Meru}, F., {Galvagni}, M., {Olczak}, C., Feb. 2013. {From Dust
  to Planetesimals: An Improved Model for Collisional Growth in Protoplanetary
  Disks}. \apj 764, 146.

\bibitem[{{Gundlach} and {Blum}(2015)}]{2015ApJ...798...34G}
{Gundlach}, B., {Blum}, J., Jan. 2015. {The Stickiness of Micrometer-sized
  Water-ice Particles}. \apj 798, 34.

\bibitem[{{G{\"u}ttler} et~al.(2010){G{\"u}ttler}, {Blum}, {Zsom}, {Ormel}, and
  {Dullemond}}]{2010A&A...513A..56G}
{G{\"u}ttler}, C., {Blum}, J., {Zsom}, A., {Ormel}, C.~W., {Dullemond}, C.~P.,
  Apr. 2010. {The outcome of protoplanetary dust growth: pebbles, boulders, or
  planetesimals?. I. Mapping the zoo of laboratory collision experiments}. \aap
  513, A56.

\bibitem[{{Haisch} et~al.(2001){Haisch}, {Lada}, and
  {Lada}}]{2001ApJ...553L.153H}
{Haisch}, Jr., K.~E., {Lada}, E.~A., {Lada}, C.~J., Jun. 2001. {Disk
  Frequencies and Lifetimes in Young Clusters}. \apjl 553, L153--L156.

\bibitem[{{Johansen} and {Klahr}(2005)}]{2005ApJ...634.1353J}
{Johansen}, A., {Klahr}, H., Dec. 2005. {Dust Diffusion in Protoplanetary Disks
  by Magnetorotational Turbulence}. \apj 634, 1353--1371.

\bibitem[{{Johansen} et~al.(2007){Johansen}, {Oishi}, {Mac Low}, {Klahr},
  {Henning}, and {Youdin}}]{2007Natur.448.1022J}
{Johansen}, A., {Oishi}, J.~S., {Mac Low}, M.-M., {Klahr}, H., {Henning}, T.,
  {Youdin}, A., Aug. 2007. {Rapid planetesimal formation in turbulent
  circumstellar disks}. \nat 448, 1022--1025.

\bibitem[{{Kataoka} et~al.(2013){Kataoka}, {Tanaka}, {Okuzumi}, and
  {Wada}}]{2013A&A...557L...4K}
{Kataoka}, A., {Tanaka}, H., {Okuzumi}, S., {Wada}, K., Sep. 2013. {Fluffy dust
  forms icy planetesimals by static compression}. \aap 557, L4.

\bibitem[{{Kretke} and {Lin}(2007)}]{2007ApJ...664L..55K}
{Kretke}, K.~A., {Lin}, D.~N.~C., Jul. 2007. {Grain Retention and Formation of
  Planetesimals near the Snow Line in MRI-driven Turbulent Protoplanetary
  Disks}. \apjl 664, L55--L58.

\bibitem[{{Kretke} et~al.(2008){Kretke}, {Lin}, and
  {Turner}}]{2008IAUS..249..293K}
{Kretke}, K.~A., {Lin}, D.~N.~C., {Turner}, N.~J., May 2008. {Planet formation
  around intermediate mass stars}. In: {Sun}, Y.-S., {Ferraz-Mello}, S.,
  {Zhou}, J.-L. (Eds.), IAU Symposium. Vol. 249 of IAU Symposium. pp. 293--300.

\bibitem[{{Leger} et~al.(1985){Leger}, {Jura}, and
  {Omont}}]{1985A&A...144..147L}
{Leger}, A., {Jura}, M., {Omont}, A., Mar. 1985. {Desorption from interstellar
  grains}. \aap 144, 147--160.

\bibitem[{{Lynden-Bell} and {Pringle}(1974)}]{1974MNRAS.168..603L}
{Lynden-Bell}, D., {Pringle}, J.~E., Sep. 1974. {The evolution of viscous discs
  and the origin of the nebular variables.} \mnras 168, 603--637.

\bibitem[{{Mathis} et~al.(1977){Mathis}, {Rumpl}, and
  {Nordsieck}}]{1977ApJ...217..425M}
{Mathis}, J.~S., {Rumpl}, W., {Nordsieck}, K.~H., Oct. 1977. {The size
  distribution of interstellar grains}. \apj 217, 425--433.

\bibitem[{{Nakamoto} and {Nakagawa}(1994)}]{1994ApJ...421..640N}
{Nakamoto}, T., {Nakagawa}, Y., Feb. 1994. {Formation, early evolution, and
  gravitational stability of protoplanetary disks}. \apj 421, 640--650.

\bibitem[{{{\"O}berg} et~al.(2011){{\"O}berg}, {Murray-Clay}, and
  {Bergin}}]{2011ApJ...743L..16O}
{{\"O}berg}, K.~I., {Murray-Clay}, R., {Bergin}, E.~A., Dec. 2011. {The Effects
  of Snowlines on C/O in Planetary Atmospheres}. \apjl 743, L16.

\bibitem[{{Okuzumi} et~al.(2016){Okuzumi}, {Momose}, {Sirono}, {Kobayashi}, and
  {Tanaka}}]{2015arXiv151003556O}
{Okuzumi}, S., {Momose}, M., {Sirono}, S.-i., {Kobayashi}, H., {Tanaka}, H.,
  Apr. 2016. {Sintering-induced Dust Ring Formation in Protoplanetary Disks:
  Application to the HL Tau Disk}. \apj 821, 82.

\bibitem[{{Okuzumi} et~al.(2009){Okuzumi}, {Tanaka}, and
  {Sakagami}}]{2009ApJ...707.1247O}
{Okuzumi}, S., {Tanaka}, H., {Sakagami}, M.-a., Dec. 2009. {Numerical Modeling
  of the Coagulation and Porosity Evolution of Dust Aggregates}. \apj 707,
  1247--1263.

\bibitem[{{Pani{\'c}} and {Min}(2017)}]{2017MNRAS.tmp..115P}
{Pani{\'c}}, O., {Min}, M., Jan. 2017. {Effects of disc midplane evolution on
  CO snowline location}. \mnras.

\bibitem[{{Pavlyuchenkov} and {Dullemond}(2007)}]{2007A&A...471..833P}
{Pavlyuchenkov}, Y., {Dullemond}, C.~P., Sep. 2007. {Dust crystallinity in
  protoplanetary disks: the effect of diffusion/viscosity ratio}. \aap 471,
  833--840.

\bibitem[{{Piso} et~al.(2015){Piso}, {{\"O}berg}, {Birnstiel}, and
  {Murray-Clay}}]{2015ApJ...815..109P}
{Piso}, A.-M.~A., {{\"O}berg}, K.~I., {Birnstiel}, T., {Murray-Clay}, R.~A.,
  Dec. 2015. {C/O and Snowline Locations in Protoplanetary Disks: The Effect of
  Radial Drift and Viscous Gas Accretion}. \apj 815, 109.

\bibitem[{{Qi} et~al.(2015){Qi}, {{\"O}berg}, {Andrews}, {Wilner}, {Bergin},
  {Hughes}, {Hogherheijde}, and {D'Alessio}}]{2015ApJ...813..128Q}
{Qi}, C., {{\"O}berg}, K.~I., {Andrews}, S.~M., {Wilner}, D.~J., {Bergin},
  E.~A., {Hughes}, A.~M., {Hogherheijde}, M., {D'Alessio}, P., Nov. 2015.
  {Chemical Imaging of the CO Snow Line in the HD 163296 Disk}. \apj 813, 128.

\bibitem[{{Ros} and {Johansen}(2013)}]{2013A&A...552A.137R}
{Ros}, K., {Johansen}, A., Apr. 2013. {Ice condensation as a planet formation
  mechanism}. \aap 552, A137.

\bibitem[{{Schr{\"a}pler} and {Henning}(2004)}]{2004ApJ...614..960S}
{Schr{\"a}pler}, R., {Henning}, T., Oct. 2004. {Dust Diffusion, Sedimentation,
  and Gravitational Instabilities in Protoplanetary Disks}. \apj 614, 960--978.

\bibitem[{{Shakura} and {Sunyaev}(1973)}]{1973A&A....24..337S}
{Shakura}, N.~I., {Sunyaev}, R.~A., 1973. {Black holes in binary systems.
  Observational appearance.} \aap 24, 337--355.

\bibitem[{{Sirono}(2011)}]{2011ApJ...735..131S}
{Sirono}, S.-i., Jul. 2011. {The Sintering Region of Icy Dust Aggregates in a
  Protoplanetary Nebula}. \apj 735, 131.

\bibitem[{{Stevenson} and {Lunine}(1988)}]{1988Icar...75..146S}
{Stevenson}, D.~J., {Lunine}, J.~I., Jul. 1988. {Rapid formation of Jupiter by
  diffuse redistribution of water vapor in the solar nebula}. \icarus 75,
  146--155.

\bibitem[{{Wada} et~al.(2009){Wada}, {Tanaka}, {Suyama}, {Kimura}, and
  {Yamamoto}}]{2009ApJ...702.1490W}
{Wada}, K., {Tanaka}, H., {Suyama}, T., {Kimura}, H., {Yamamoto}, T., Sep.
  2009. {Collisional Growth Conditions for Dust Aggregates}. \apj 702,
  1490--1501.

\bibitem[{{Weidenschilling}(1977)}]{1977MNRAS.180...57W}
{Weidenschilling}, S.~J., Jul. 1977. {Aerodynamics of solid bodies in the solar
  nebula}. \mnras 180, 57--70.

\bibitem[{{Windmark} et~al.(2012){Windmark}, {Birnstiel}, {Ormel}, and
  {Dullemond}}]{2012A&A...544L..16W}
{Windmark}, F., {Birnstiel}, T., {Ormel}, C.~W., {Dullemond}, C.~P., Aug. 2012.
  {Breaking through: The effects of a velocity distribution on barriers to dust
  growth}. \aap 544, L16.

\bibitem[{{Yamamoto} et~al.(1983){Yamamoto}, {Nakagawa}, and
  {Fukui}}]{1983A&A...122..171Y}
{Yamamoto}, T., {Nakagawa}, N., {Fukui}, Y., Jun. 1983. {The chemical
  composition and thermal history of the ice of a cometary nucleus}. \aap 122,
  171--176.

\bibitem[{{Youdin} and {Goodman}(2005)}]{2005ApJ...620..459Y}
{Youdin}, A.~N., {Goodman}, J., Feb. 2005. {Streaming Instabilities in
  Protoplanetary Disks}. \apj 620, 459--469.

\bibitem[{{Youdin} and {Shu}(2002)}]{2002ApJ...580..494Y}
{Youdin}, A.~N., {Shu}, F.~H., Nov. 2002. {Planetesimal Formation by
  Gravitational Instability}. \apj 580, 494--505.

\end{thebibliography}

  \appendix

\section{Modified Podolak algorithm for $Q$}

\citet{2008A&A...480..859B} described, in their appendix A, the difficulties of
numerically solving the discrete Smoluchowski equation for coagulation on a
logarithmically spaced mass grid. They present a method for circumventing these
problems. This method has to be modified for the $Q$ parameter used in this
work.

We will shortly repeat the formalism of \citet{2008A&A...480..859B}. For further
details we refer to that paper.

By coagulating two particles on a logarithmic mass grid, the resulting
particle's mass will not be exactly on a grid point. Therefore, its mass has to
be split between two adjacent grid points. Let us assume the collision rate
$R_{ij}$ of particles with masses $m_i$ and $m_j$ is given by
\begin{equation}
  R_{ij} = N_i N_j K_{ij},
\end{equation}
where $N_i$ is the number density of particle species $i,$ and $K_{ij}$ is the
coagulation Kernel of mass bins $i$ and $j$. The resulting particle mass of
these collisions is $m = m_i + m_j$ , which lies in between the two mass grid
points $m_m$ and $m_n$ such that $m_m < m < m_n$. The change in number density
of mass bin $k$ is then given by
\begin{equation}
  \dot{N}_k = \frac{1}{2} \sum\limits_{ij} R_{ij} C_{ijk} - \sum\limits_i
R_{ik},
\end{equation}
with
\begin{equation}
  C_{ijk} = \begin{cases}
              \epsilon & \text{if } m_k \text{ is the largest grid point with }
m_k < m \\
              1 - \epsilon & \text{if } m_k \text{ is the smallest grid point
with }
m_k > m \\
              0 & \text{else}
            \end{cases}
,\end{equation}
and
\begin{equation}
  \epsilon = \frac{m_n - m}{m_n - m_m}.
\end{equation}
This guarantees that the collision rates are split between two adjacent
grid points with strict mass conservation, which can be easily calculated.

Another problem arises if particles coagulate, whose masses differ by more than
15 orders of magnitude. Standard double precision variables have accuracies up
to 15 digits. In that case, for a computer using double precision variables, a
sticking collision of two particles with masses $m_i$ and $m_j$ would lead to a
particle with mass $m = m_i + m_j = m_j$ when $m_j$ is the larger particle,
meaning many collisions of very small particles with a very large particle would
lead to no growth at all, which is solely a numerical effect.

\citet{2008A&A...480..859B} avoided this by re-sorting sums. They came to the
following equation:
\begin{equation}
  \dot{N}_k = \sum\limits_{ij} N_i N_j K_{ij} M_{ijk},
\end{equation}
where the matrix $M$ is given by
\begin{equation}
  \begin{split}
    M_{ijk} = & \delta_{ij} C_{ijk} + C_{ijk} \Theta \left( k - i - \frac{3}{2}
\right) \Theta \left( i - j - \frac{1}{2} \right) + D_{ji} \delta_{ik} \\
              & + E_{j,i+1} \delta_{i,k-1} \Theta \left( k - j - \frac{3}{2}
\right),
  \end{split}
\end{equation}
where $\delta$ is the Kronecker delta and $\Theta$ the Heaviside step function.
The matrices $D$ and $E$ are given by
\begin{equation}
  D_{jk} = \begin{cases}
             - \frac{m_j}{m_{k+1} - m_{k}} & l \leq k + 1 - c_e \\
             1 & \text{else}
           \end{cases}
,\end{equation}
and
\begin{equation}
  E_{jk} = \begin{cases}
             - \frac{m_j}{m_k - m_{k-1}} & j \leq k - c_e \\
             1 - \frac{m_j + m_{k-1} - m_k}{m_{k+1}-m_k} & j > k - c_e\ \wedge\
m_{k-1} + m_j \leq m_{k+1} \\
             0 & \text{else}.
           \end{cases}
\end{equation}
The integer number $c_e$ is defined in a way that the inequality,
\begin{equation}
  m_{k-1} + m_i < m_k
,\end{equation}
is satisfied for every $i \leq k - c_e$. In that way, some equal sized terms
cancel out in the sums and inaccuracies resulting from the limited accuracy of
the double precision variables can be avoided.
Unfortunately, by introducing the $Q$ parameter, these terms do not exactly
cancel out in the coagulation equation for $NQ_i \equiv N_i \cdot Q_i$. We have
to modify the equations here to get to
\begin{equation}
  \dot{NQ_k} = \sum\limits_{ij} N_i N_j K_{ij} \tilde{M}_{ijk},
\end{equation}
with the matrix $\tilde{M}$ that can be expressed in terms of $M$ as
\begin{equation}
  \tilde{M}_{ijk} = Q^\mathrm{new}_{ij} \left( M_{ijk} - D_{ji} \delta_{ik} +
\hat{D}_{ji} \delta_{ik} \right),
\end{equation}
and
\begin{equation}
  \hat{D}_{ji} = \begin{cases}
                   1 - \frac{Q_i}{Q^\mathrm{new}_{ij} - \frac{m_j}{m_{i+1}-m_i}}
& j \leq i + 1 - c_e \\
                   -\frac{Q_i}{Q^\mathrm{new}_{ij}} & \text{else}.
                 \end{cases}
\end{equation}
$Q^\mathrm{new}_{ij}$ is the new $Q$-value of the particle that resulted from
the collision of particle $i$ with particle $j$.

\end{document}